\documentclass[a4paper, UKenglish, cleveref, autoref, thm-restate, numberwithinsect]{lipics-v2021}

\nolinenumbers
\usepackage[disable]{todonotes}

\usepackage{physics}

\usepackage[appendix=inline]{apxproof}\usepackage[inline]{optional}

\newtheoremrep{theorem}{Theorem}[section]
\newtheoremrep{lemma}[theorem]{Lemma}
\newtheoremrep{corollary}[theorem]{Corollary}

\usepackage{sidecap}
\usepackage{gensymb}
\usepackage{cite}
\usepackage{inconsolata}

\usepackage{mathtools} 
\usepackage{extarrows}

\usepackage[normalem]{ulem}

\newcommand{\blank}{\ensuremath{\llcorner\negthinspace\lrcorner}}

\hypersetup{
    colorlinks=true,
    linkcolor=blue,
    filecolor=magenta,      
    urlcolor=blue,
    citecolor=blue,
    pdftitle={Overleaf Example},
    pdfpagemode=FullScreen,
}
\title{Simulating 3-symbol Turing machines with SIMD||DNA}

\titlerunning{Simulating 3-symbol Turing machines with SIMD||DNA}

\author
{David Doty}
{University of California, Davis, USA \and \url{https://web.cs.ucdavis.edu/~doty/}}
{doty@ucdavis.edu}
{https://orcid.org/0000-0002-3922-172X}
{}

\author
{Aaron Ong}
{University of California, Davis, USA}
{aabong@ucdavis.edu}
{}
{}

\authorrunning{D. Doty, A. Ong}

\Copyright{David Doty, Aaron Ong}

\ccsdesc[100]{Theory of computation~Models of computation}

\keywords{DNA storage, strand displacement, parallel computation} 

\category{} 

\supplement{
    simulator:
    \url{https://github.com/UC-Davis-molecular-computing/simd-dna}
} 

\acknowledgements{The authors were supported by NSF grants 1900931 and 1844976.}

\begin{document}

\maketitle

\begin{abstract}
    SIMD||DNA~\cite{wang2019simd} is a model of DNA strand displacement allowing parallel in-memory computation on DNA storage.
    We show how to simulate an arbitrary 3-symbol space-bounded Turing machine with a SIMD||DNA program, 
    giving a more direct and efficient route to general-purpose information manipulation on DNA storage than the Rule 110 simulation of Wang, Chalk, and Soloveichik~\cite{wang2019simd}.
    We also develop software~\cite{simd_simulator} that can simulate SIMD||DNA programs and produce SVG figures.
\end{abstract}


\section{Introduction}
\label{sec:intro}

\begin{figure}[ht]
    \centering
    \includegraphics[width=13cm]{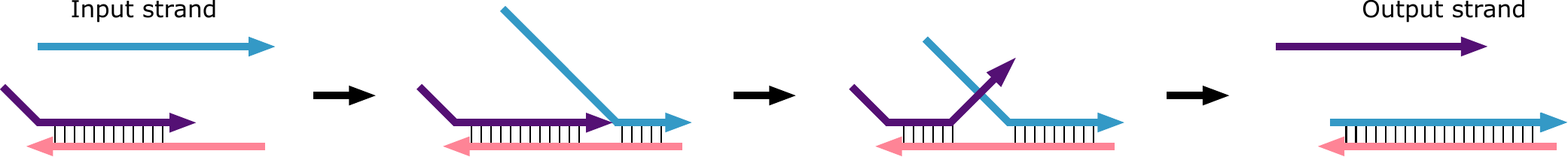}
    \caption{DNA strand displacement (see~\cite{seelig2006enzyme} for more details). An input DNA strand (turquoise) binds to the short toehold region of a complementary strand (pink) and displaces the output strand (purple).
    The toehold region is so-called because, although too short to bind stably, 
    it allows temporary binding of the input, giving it a ``foot in the door'' to begin the displacement process.}
    \label{fig:dsd}
\end{figure}

DNA storage typically encodes information in the choice of DNA sequences~\cite{church2012next, organick2018random, bornholt2016dna},
so that reading and writing require expensive sequencing (reading DNA) and synthesis (writing DNA) steps.
An alternative ``nicked storage'' scheme of Tabatabaei et al.~\cite{tabatabaei2020dna} uses a single long strand called a \emph{register}, 
with a fixed sequence.
Information is stored in the choice of short complementary strands to bind to the register.
This gives the potential to process the stored information using \emph{DNA strand displacement} (see \cref{fig:dsd}), 
which reconfigures which DNA strands are bound, without changing their sequences.
Thus manipulation of the stored information (i.e., computation) can potentially be done \emph{in vitro} with simpler lab steps than DNA sequencing or synthesis.

The SIMD||DNA model of Wang, Chalk, and Soloveichik~\cite{wang2019simd} is an abstract model of such a system.
It allows parallel in-memory computation on several copies of the register;
each register may store different data.
In the experimental implementation,
each register strand is attached to a magnetic bead, enabling \emph{elution}:
washing away strands not bound to a register, 
while keeping the registers (and their bound strands) in the solution due to their attachment to the bead.
This motivates the ``multi-stage'' SIMD||DNA model of DNA strand displacement, which at a high level works as follows.
Each stage is called an \emph{instruction},
consisting of a set of strands to add to the solution.
It is assumed that strand displacement reactions proceed until the solution reaches equilibrium,
at which point all strands and complexes not attached to a register are washed away.
The strands for the next instruction are then added.
A key aspect of the model is that the wash step can constrain what strand displacement reactions are possible afterward, compared to ``one-pot'' strand displacement schemes that mix all strands from the start.
This gives the SIMD||DNA model potentially more power than one-pot DNA strand displacement.
Wang, Chalk, and Soloveichik~\cite{wang2019simd} showed SIMD||DNA programs for binary counting and simulating cellular automata Rule 110,
and Chen, Solanki, and Riedel~\cite{chen2021parallel} showed SIMD||DNA programs for sorting, shifting and searching in parallel.
See~\cref{sec:model} for a formal definition and~\cite{wang2019simd} for more details and motivation for the model.

A major theoretical result of~\cite{wang2019simd} is a SIMD||DNA program that simulates a space-bounded version of cellular automata Rule 110.
When space is unbounded,
Rule 110 is known to be efficiently Turing universal, i.e.,
able to simulate any single-tape Turing machine~\cite{cook2004universality}
with only a polynomial-time slowdown~\cite{neary2006p}, though by an awkward indirect construction and encoding with very large constant factors.
We show how to simulate an arbitrary 3-symbol space-bounded single-tape Turing machine \emph{directly} with a SIMD||DNA program.
Since custom manipulation of bits is much easier to program in a Turing machine than Rule 110,
this gives a more direct, efficient, and conceptually simple method of general-purpose information processing on nicked DNA storage.
Although we have not worked out the details,
it seems likely that the construction can be extended straightforwardly to Turing machines with alphabet sizes larger than 3.
However, it is straightforward to simulate a larger-alphabet Turing machine $M$ with a 3-symbol Turing machine $S$, for example representing each of 16 non-blank symbols of $M$ by 4 consecutive bits of $S$.

Our construction was designed and tested using software we developed~\cite{simd_simulator} for simulating the SIMD||DNA model.
It is able to take a description of an arbitrary SIMD||DNA program: 
a list of instructions,
where each instruction is a set of DNA strands to add.
It produces figures indicating visually how the steps work, 
both with text printed on the command line (for quickly testing ideas)
and SVG figures,
such as most of those in this paper.


\section{Model}
\label{sec:model}

In this section we define the model of SIMD||DNA~\cite{wang2019simd}.

\begin{figure}[ht]
    \centering
    \includegraphics[width=0.89\textwidth]{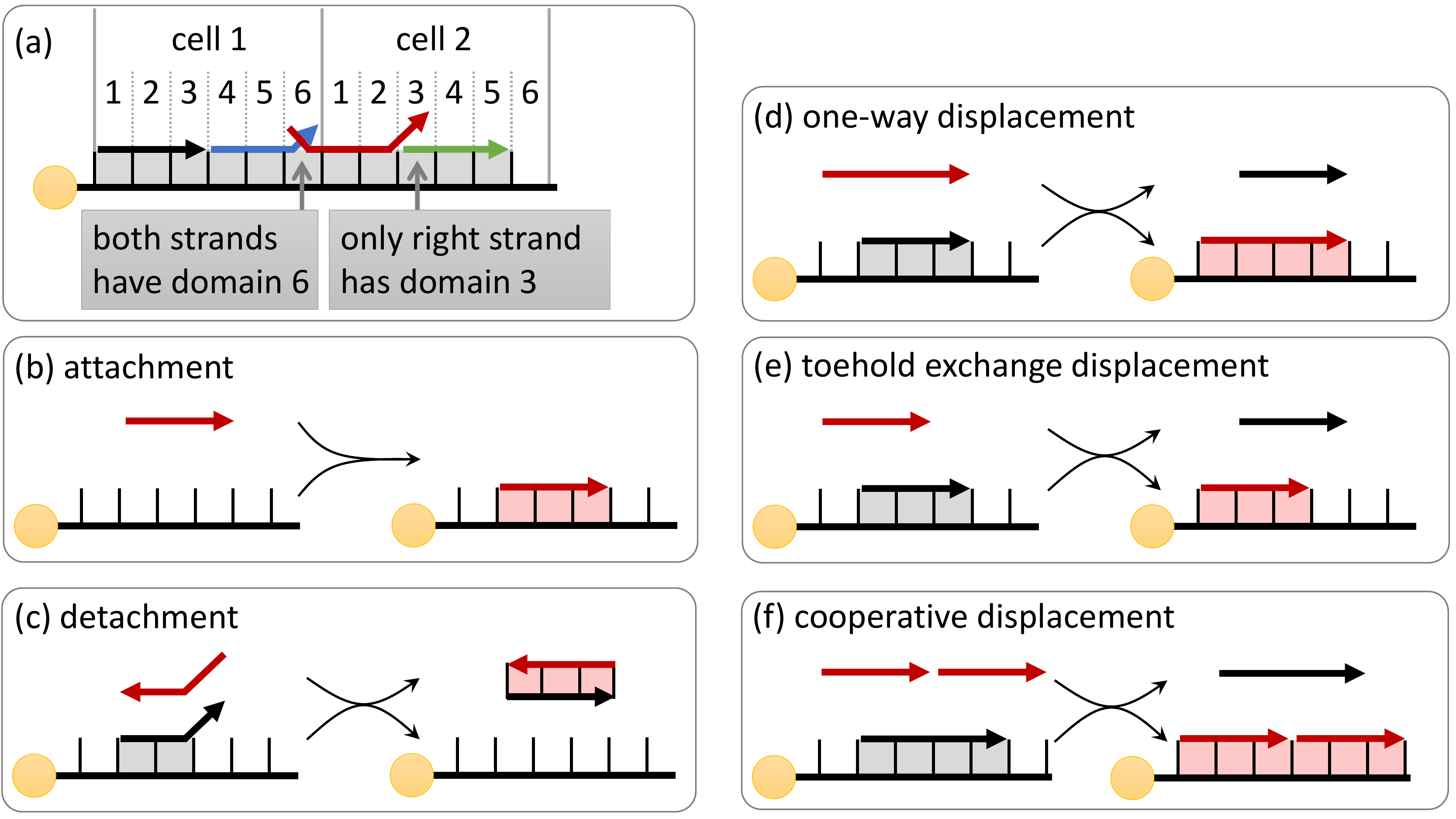}
    \caption{ \footnotesize
    Notational conventions and reactions in the SIMD||DNA model.
    The register strand is on the bottom in each subfigure, with a yellow round ``magnetic bead'' depicted on the left (not depicted in subsequent figures).
    Lightly shaded gray or pink regions denote bonds (double-stranded regions),
    but later figures omit this and simply draw a forward strand (one with $5'$ end on left and $3'$ end on right) immediately above the domains to which it is bound on the register strand.
\textbf{(a)}
    Conventions for domain names of strands.
    Domains are numbered $1,\ldots,d$ within each cell; $d=6$ in~\cref{fig:simd-model}(a) and $d=18$ in subsequent figures.
    The register strand has the starred versions of these domains.
    If a top strand is horizontal over domain $i$, it has domain $i$.
    If it is diagonal over the whole domain, it has an unlabelled domain distinct from all register domains (used as a toehold overhang for detachment, see subfigure (c)).
    If two strands both partially cover a domain then they both have that domain.
\textbf{(b)}
    A forward instruction strand can attach if at least two complementary consecutive domains are unbound on the register. 
\textbf{(c)}
    Reverse instruction strands can bind to toehold overhangs on forward bound strands to detach them from the register.
    The fact that the (unlabelled) toeholds are complementary is indicated by a diagonal bend in the reverse strand matching.
\textbf{(d)}
    Forward instruction strands can do toehold-mediated strand displacement, one-way if the displacing strand contains all the domains of the displaced strand.
\textbf{(e)}
    If the displacing strand is missing the last domain of the displaced, displacement can also happen, known as \emph{toehold exchange}.
    This is often called ``reversible'' since it conserves the number of bound domains, but in the SIMD||DNA model,
    instruction strands are added in large excess over registers,
    making it effectively irreversible due to the entropic bias toward binding the instruction strand.
    Thus it is depicted with irreversible arrows in the figure.
\textbf{(f)}
    Two forward strands can cooperate to displace a single bound top strand,
    even if neither has enough domains to displace on its own.
    }
    \label{fig:simd-model}
\end{figure}

See~\cref{fig:simd-model} for notational conventions in the SIMD||DNA model and an explanation of the basic strand displacement reactions.
The register strand is on the bottom in each sub-figure, with a yellow round ``magnetic bead'' depicted on the left (bead not depicted in subsequent figures).
A DNA strand has an orientation, with one end called the $5'$ end and the other called the $3'$ end;
by convention strands are drawn as arrows with the arrowhead on the $3'$ end.
The register strand has its $3'$ end on the left and $5'$ end on the right.
The model allows multiple registers to be present in solution at once, each possibly configured differently.
However, it is assumed that register strands are sufficiently dilute that they do not interact with each other or with strands that have been displaced from other registers.
Thus all figures depict only a single register and instruction strands that interact with it.

The register strand is divided into \emph{cells}, which are further divided into \emph{domains}.
Each domain can be thought of as a fixed-length DNA sequence with relatively weak binding (e.g., 5-7 bases).
A strand is stably attached to the register strand only if it is bound by at least two domains,
but one domain is sufficiently long to act as a ``toehold'' to help initiate strand displacement
(\cref{fig:simd-model}(c-f)).
Within a cell with $d$ domains, each domain is unique and assumed to be named $1,2,\ldots,d$.
The register strand has the starred version of these domains, e.g., $1^*,2^*,3^*,4^*,5^*,6^*,1^*,2^*,3^*,4^*,5^*,6^*$ reading from the register's $3'$ to $5'$ end (left to right) in~\cref{fig:simd-model}(a).
All cells have the same ordered list of domains,
so for example in~\cref{fig:simd-model}(a),
domain 5 in cell 1 is the same DNA sequence as domain 5 in cell 2.

An \emph{instruction} is a set of strands that are added to the solution at once.
\cref{fig:simd-model}(b-f) shows the various reactions that these strands might conduct to change the configuration of strands attached to the register.
Multiple reactions can occur in a cascade in a single instruction.\footnote{
    See for example instruction 39 in~\cref{fig:transition-left-cell-has-1-first-half}.
    In the right cell, 
    an orange instruction strand displaces an orange strand bound to the register via toehold exchange.
    This opens a toehold for a blue instruction strand to displace the bound blue strand, 
    resulting in the configuration shown at the beginning of instruction 41.
}
In particular, the model is nondeterministic, and in general multiple reactions might be possible.
It is the job of the system designer to ensure that only one final configuration can result no matter the order of reactions.
Instruction strands can either be \emph{forward} ($3'$ arrow on right)
or \emph{reverse} ($3'$ arrow on left).
Forward instruction strands can do attachment and displacement reactions
(\cref{fig:simd-model}(b,d-f))
and reverse instruction strands can detach forward strands previously bound to the register
(\cref{fig:simd-model}(c)).

Crucially, instruction strands are added in large excess over the register strands.
Thus even the toehold exchange displacement,
which is often considered reversible due to being enthalpically balanced (same number of domains bound before and after), 
is actually irreversible in the SIMD||DNA model,
due to the entropic bias toward binding the new instruction strand with much larger concentration than the strand it displaces.


The notation of~\cite{wang2019simd} uses dashed lines for reverse strands used for detachment, 
as a visual reminder that they do not bind to the register.
We leave reverse strands as solid lines and rely on the $3'$ arrow to denote that the strand is reversed.
We reserve the dashed line notation for later figures to depict \emph{inert} instruction strands:
instruction strands that are shown above the register where they would bind if possible,
but where no reaction allows them to do so in the current configuration.
We also have a slightly different notation for strands with domains mismatching the register:
in~\cite{wang2019simd},
these are depicted by writing an explicit domain name.
In our convention,
the drawing of that part of the strand as diagonal and lying entirely above the register domain indicates that the top strand domain and register domain are not complementary
(\cref{fig:simd-model}(a), cell 2, domain 3).
To denote that two adjacent top strands share the same domain, both of which can bind to the register (so they dynamically compete with strand displacement),
we draw both strands partially horizontal over the domain, and partially diagonal
(\cref{fig:simd-model}(a), cell 1, domain 6).

Although these rules allow for nondeterministically competing reactions,
our construction is deterministic in the sense that there is only one sequence of reactions possible in any instruction step.

After instruction strands are added and the described reactions go to completion,
the \emph{wash} step removes all strands not bound to the register.
This includes excess instruction strands that never reacted,
as well as strands that were displaced or complexes formed in a detachment reaction.
\section{Simulation of Turing machine in SIMD||DNA}
\label{sec:simulation}

In this section we describe how to simulate an arbitrary 3-symbol single-tape Turing machine with SIMD||DNA instructions.

\subsection{High-level overview of construction}
\begin{figure}[ht]
    \centering
    \includegraphics[width=8cm]{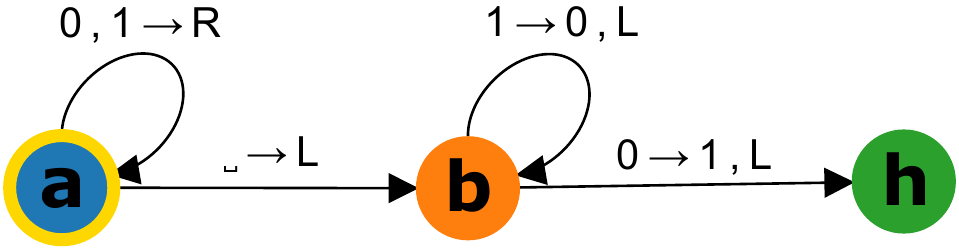}
    \caption{
    Turing machine (start state $a$) that increments an integer represented in binary, with the least significant bit on the right.
    This example is simulated in all subsequent figures.
    }
    \label{fig:binary_tm}
\end{figure}

Since the SIMD||DNA model as defined has no mechanism to grow the register strand,
it can only simulate a fixed-space-bound Turing machine (a.k.a., linear-bounded automaton),
which starts with $s$ total tape cells and never moves the tape head off of them.
A 3-symbol, space-$s$ Turing machine has three tape symbols: $0,1,\blank$.
The binary input $x \in \{0,1\}^{< s}$ is represented by string $x \blank^{s-|x|}$ on the tape in the initial configuration,
i.e. $x$ padded with enough blank symbols to make $s$ total tape cells.
We use as a running example the 5-transition Turing machine in~\cref{fig:binary_tm},
which increments a binary number.

Each cell of the register represents a tape cell of the Turing machine. 
If the Turing machine has $t$ total transitions,
then each cell uses
$d = 2t + 8$ domains.

For each Turing machine,
there is a fixed sequence of instructions that, after executing,
will update the register to represent the next configuration of the Turing machine.

The cell with the tape head is the only cell with uncovered register domains.
Which domains are uncovered
(known as a \emph{transition region})
represents both the current state of the Turing machine and the symbol written on that tape cell.
For all other cells, a disjoint region 
(the \emph{symbol region})
represents the symbol on that cell through its pattern of nicks.
On the cell with the tape head, the symbol region has no nicks (and represents no symbol) since it is covered by a longer 8-domain strand.

\subsection{Representation of Turing machine tape cell as a register cell}
In the SIMD||DNA representation of a Turing machine, each register cell represents a single Turing machine tape cell. 
We represent each Turing machine with tape alphabet $\Gamma = \{0,1,\blank\}$, state set $Q$, and halt state $h$,
as a set of \emph{transitions},
where each transition
$(q,b) \to (r,c,m)$
means that if the Turing machine is in state $q \in Q \setminus \{h\}$ reading symbol $b \in \Gamma$,
it changes to state $r$, writes symbol $c$, and moves one cell by $m \in \{L,R\}$ (left or right).
Since the Turing machine is deterministic,
for each state-symbol pair,
there is at most one transition with that pair on the left.
(But some such pairs could be undefined, e.g., there is no $(b,\blank) \to  \ldots$ transition in~\cref{fig:binary_tm}.)

\paragraph*{Representation of tape cell with tape head}

\begin{figure}[ht]
    \centering
    \includegraphics[width=0.5\textwidth]{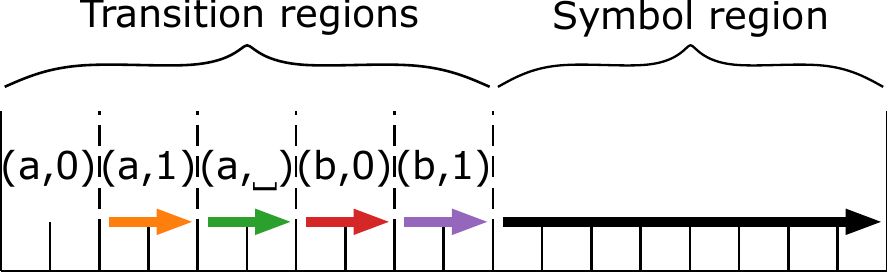}
    \caption{
    A SIMD||DNA cell where the tape head is presently located. 
    The $(a,0)$ region is fully exposed, indicating that the Turing machine is in state $a$ and that the cell contains the symbol $0$. The other transition regions are fully covered,
    and the symbol region (rightmost 8 domains of the cell) is covered by a single long strand, not encoding any symbol (which is encoded by the uncovered transition region). }
    \label{fig:tape_cell}
\end{figure}

We take every state-symbol pair $(q, \sigma) \in (Q \setminus \{h\}) \times \Gamma$ (each possible left side of a transition) and represent each as two consecutive domains in a SIMD||DNA register cell. 
See~\cref{fig:tape_cell}.
Recall the binary incrementing Turing machine of~\cref{fig:binary_tm}.
It has five transitions: 
$(a, 0) \to (a,0,R),\quad
(a, 1) \to (a,1,R),\quad
(a, \blank) \to (b,\blank,L),\quad
(b, 0) \to (1,h,L),\quad
(b, 1) \to (0,b,L)$.
We call the pair on the left the \emph{transition input}.
Each of the given transition inputs is represented in the SIMD||DNA cell using two domains, requiring ten domains total for our example. 
Since each register cell represents a cell in $M$, we must denote the presence of the tape head on one of the cells. If the tape head is present on a given cell and if the current Turing machine configuration has a valid transition, then the two domains that represent that transition will have no top strand attached to them, leaving them exposed. 
For example, if the tape head is on a cell with the $0$ symbol, and the Turing machine is currently in state $a$, then the region that represents $(a, 0)$ in that cell will be exposed to serve as a toehold for strand displacement. 
The other transition regions are fully covered by 2-domain strands.

\paragraph*{Representation of tape cell without tape head}
If the tape head is not present on a cell, or if no valid transitions exist for the current configuration,\footnote{
    For example, if the machine has halted; see the bottom register configuration of~\cref{fig:transition-left-cell-has-1-second-half} for a case where the state is non-halting but no valid transition exists.
}
then every transition region is covered by 2-domain strands. Eight additional domains at the rightmost part of the cell, called the \textbf{symbol region} represent the current symbol written on that cell.

\begin{figure}[ht]
    \centering
    \makebox[\textwidth][c]{\includegraphics[width=1.0\textwidth]{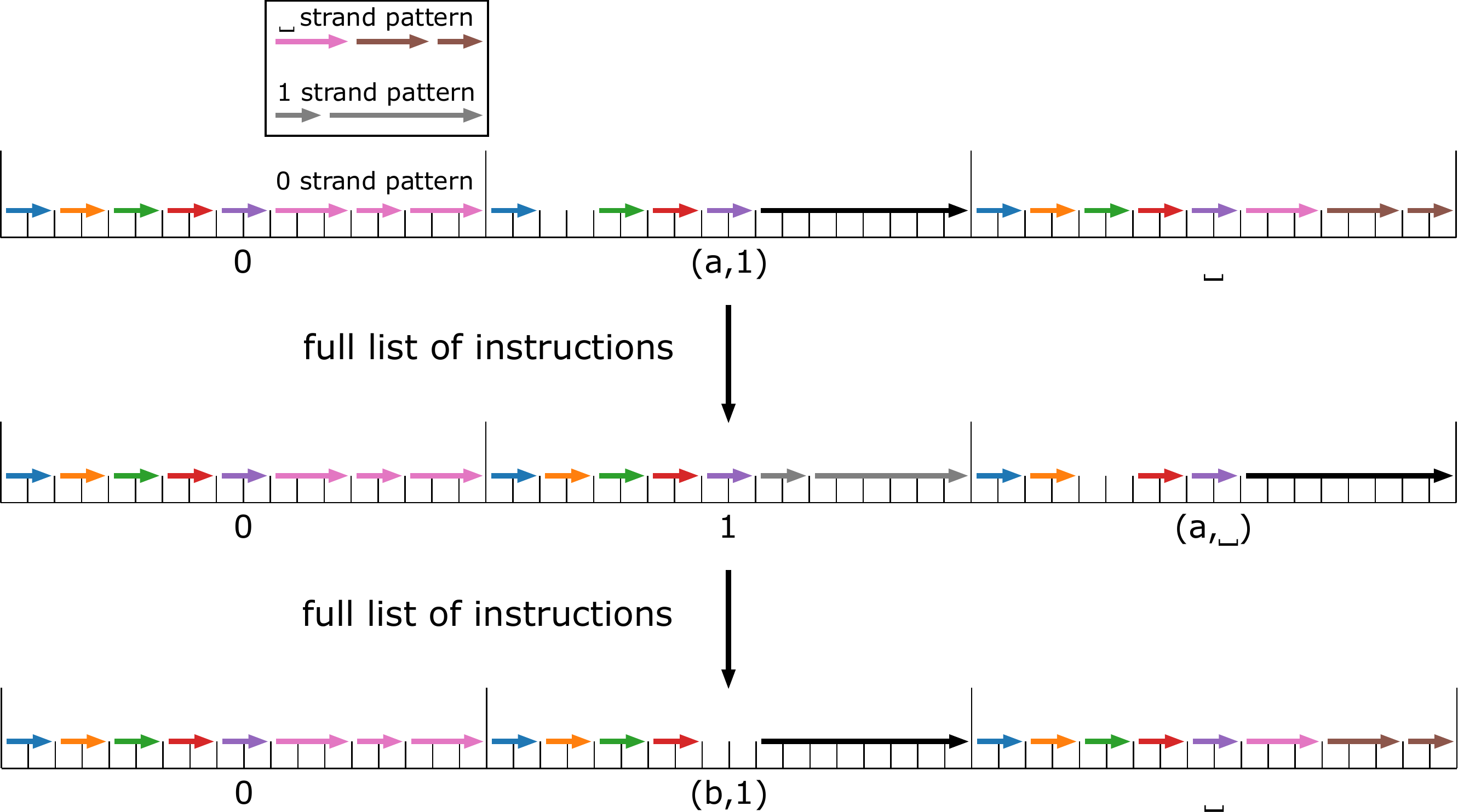}}%
    \caption{
    High-level overview of construction.
    A Turing machine register currently in state $a$, with the tape head on the second cell. The second cell contains the symbol 1. The leftmost cell contains symbol 0.
    The inset above shows encodings for 1 and \blank.
    All the transition regions are fully covered in cells lacking the tape head.
    After the full list of instructions in the SIMD||DNA program are complete,
    the register represents the Turing machine configuration with state $a$ and the tape head moved to the rightmost cell with the \blank.
    The same full list of instructions updates the register again,
    now representing the Turing machine configuration in state $b$ with the tape head back on the middle cell.
    }
    \label{fig:tm_example}
\end{figure}

\begin{figure}[ht]
    \centering
    \includegraphics[width=13.5cm]{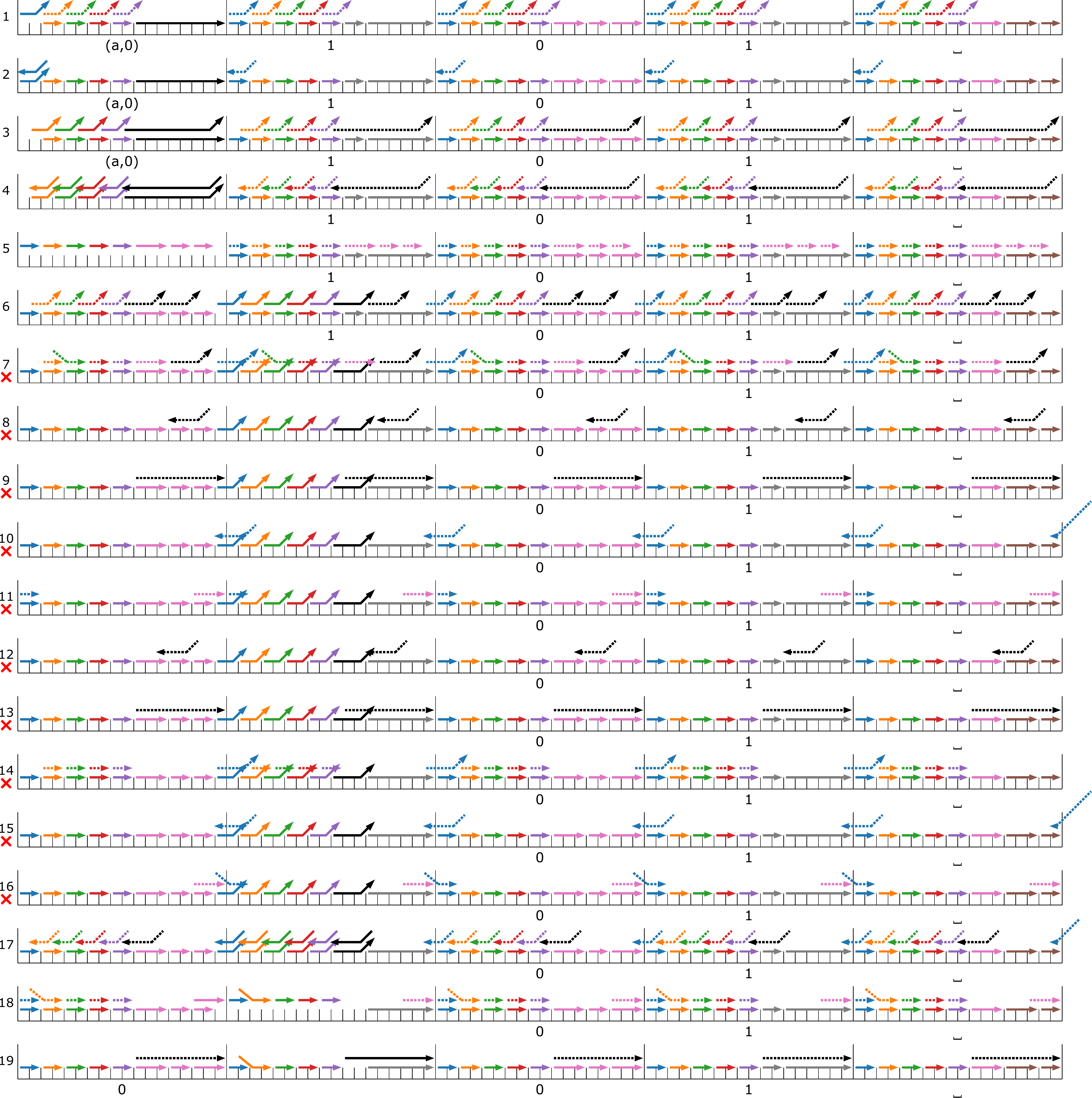}
    \caption{
    An overview of the first 19 instructions of the construction, which represent the $(a, 0) \to (a,0,R)$ transition of the Turing machine shown in \cref{fig:binary_tm}. Instructions marked with a red cross are fully inert, meant for cases not exhibited by this register.}
    \label{fig:a0_instructions}
\end{figure}

Whenever the tape head is present on a cell, the symbol region is covered by a single 8-domain strand that does not encode any symbol, since the symbol information is already encoded in the transition region with an open toehold.

\subsection{Detailed description of  SIMD||DNA instructions simulating a Turing machine}
We designed an algorithm that converts Turing machine specifications from \url{https://turingmachine.io} into SIMD||DNA representations, along with the equivalent instructions. 
Each transition $\tau_i$ has an associated sublist of instructions $L_i$, and,
not knowing which transition is applicable to the current configuration,
we simply add instruction strands in order from $L_1, L_2, \ldots$.
For $i \neq j$, to ensure that $L_j$ instructions have no effect when the current applicable transition is $\tau_i$,
we ``plug'' the open domains of other transition regions with a strand and remove the plug strand once it's time to process that transition. Because the SIMD||DNA model allows parallel computation among multiple registers in the same solution, this prevents instructions meant for one configuration from affecting registers currently not in that configuration. In the beginning, all transition regions are plugged, where the order of processing for the transitions is arbitrary.

\begin{figure}[ht]
    \centering
    \includegraphics[width=12cm]{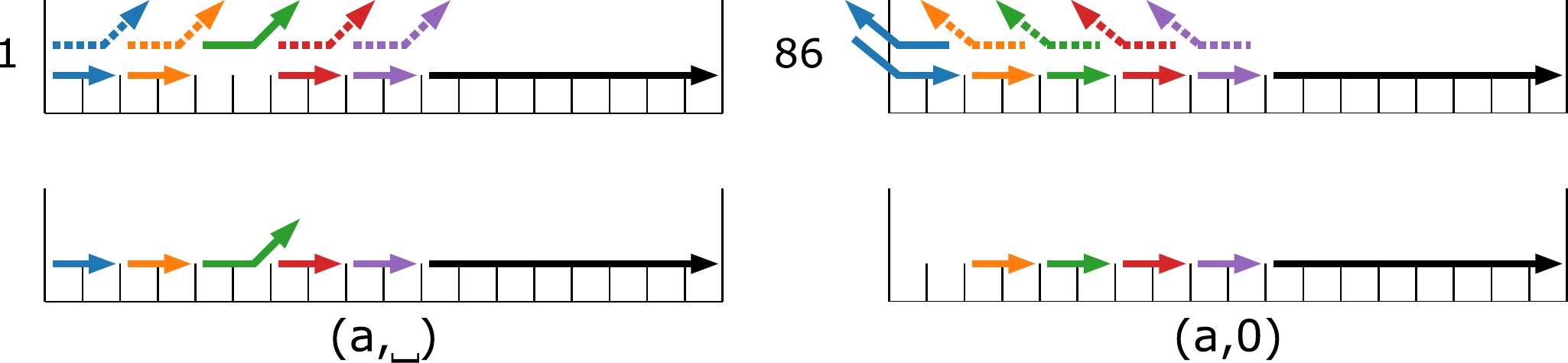}
    \caption{ \footnotesize
    On the left, the first instruction in the whole SIMD||DNA program covers the transition region representing the next applicable transition.
    The two horizontal rows of strands have the following interpretation:
    Bottom are strands bound to register,
    top are instruction strands. 
    Dashed instruction strands will not have an effect on the current cell (but to help verify correctness, they are shown above where they would bind to the register).
    On the right, the last instruction in the whole SIMD||DNA program, which removes the post-plug strands in each register. In the above example, the post-plug strand covers the $(a,0)$ transition region, indicating the cell's next Turing machine transition.
    After this, the entire register is updated to appear as a configuration similar to those in~\cref{fig:tm_example}.
    }
    \label{fig:plug_and_final}
\end{figure}

\begin{figure}[!ht]
    \centering
    \includegraphics[width=10cm]{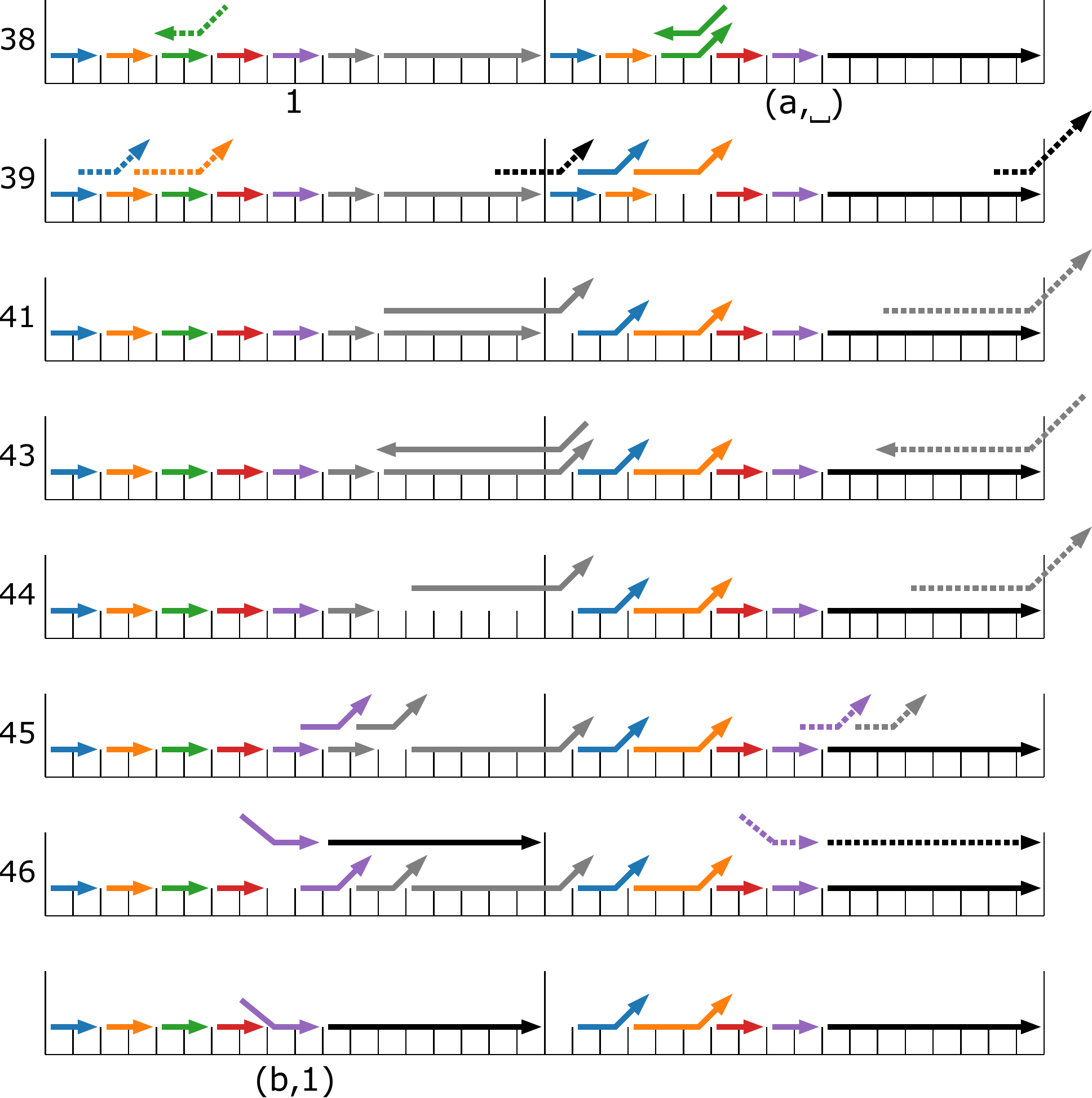}
    \caption{
    First half of instructions
    (next-cell instructions) 
    to implement transition
    $a,\blank \to b,\blank,L$, in the case that the cell to the left (where the tape head will move) has a symbol 1.
    After instruction 46, the left cell (where the tape head will move next) now encodes the next state $b$ and the symbol 1 on the new tape cell.
    Rather than show a red cross, inert instructions are instead omitted in this figure and the following ones. Inert instructions (40,42,47,48) are used when the cell to the left has a symbol 0 or \blank\ on it instead.
    \cref{fig:transition-left-cell-has-0,fig:transition-left-cell-has-blank}
    respectively show these cases.
    \cref{fig:transition-left-cell-has-1-second-half} shows instructions completing the transition by writing \blank\ over the right cell.
    In instruction 38, the transition region is first unplugged to expose the toehold. The strands in instruction 39 cascade until the left cell's symbol region, allowing the instructions to branch out depending on the tape content of the left cell. The remaining instructions process the left cell so that it encodes the next configuration. In instruction 46, a special post-plug strand whose leftmost domain is orthogonal is attached to the region that represents the next configuration; this strand will be removed once all transitions have been processed. The angled dashed strands to the right of the register indicate that no cells are present to the right of the rightmost cell; if there were another cell, then one more domain of each of these strands would be horizontal, bound to the leftmost domain of the cell to the right (just as with their solid counterparts to the left).}
    \label{fig:transition-left-cell-has-1-first-half}
\end{figure}

\begin{figure}[ht]
    \centering
    \includegraphics[width=10cm]{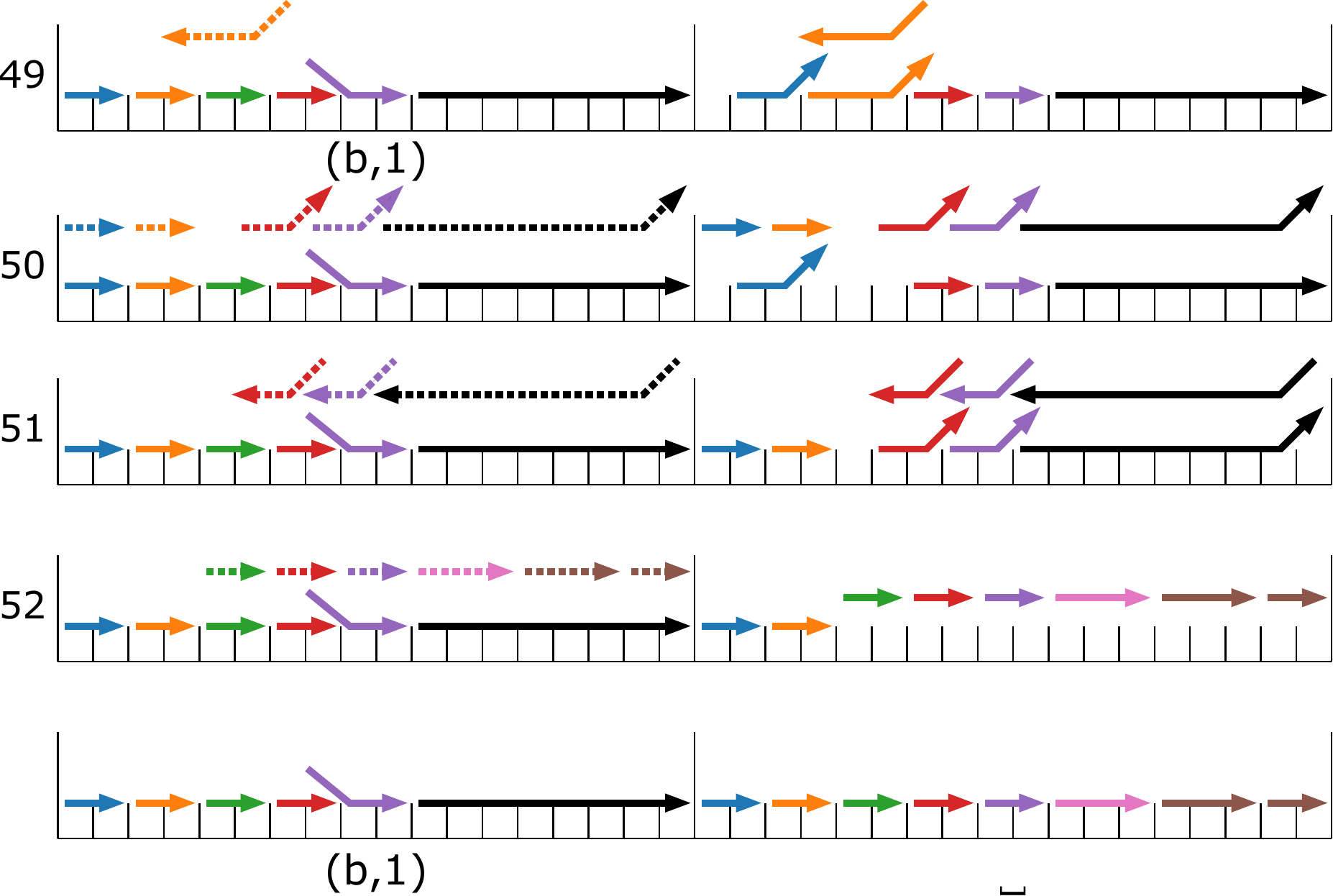}
    \caption{
    Second half (previous-cell instructions) of transition 
    $a,\blank \to b,\blank,L$ whose first 11 instructions are shown in~\cref{fig:transition-left-cell-has-1-first-half}, \cref{fig:transition-left-cell-has-0}, and~\cref{fig:transition-left-cell-has-blank}; 
    these instructions write \blank\ over the old cell (right of~\cref{fig:transition-left-cell-has-1-first-half}) where the tape head was at the start of the transition.
    Slight variations of instruction 15 write 0 or 1 instead of \blank.
    }
    \label{fig:transition-left-cell-has-1-second-half}
\end{figure}

\paragraph*{Pre-plug and post-plug strands to protect instructions for inapplicable transitions from affecting configuration}
The full list of instructions to simulate a Turing machine transition works as follows.
Recall that in the ``clean'' configurations shown in~\cref{fig:tm_example},
the only exposed register domains are on the cell representing the tape head.
The first instruction in the entire list contains \emph{pre-plug} strands for each transition region. At the end of each instruction sublist $L_i$, a \emph{post-plug strand} is also placed on the transition region that represents the Turing machine's next applicable transition. Examples of both can be seen in~\cref{fig:plug_and_final}.
These strands act like a ``chemical protecting group'' that prevents instruction sublists $L_i$ from modifying the register unless they apply to the intended transition. 
The difference between the pre-plug and post-plug strands is that pre-plug strands protect the configuration when using instructions strands \emph{before} the applicable transition, whereas post-plug strands protect the configuration when using instructions strands \emph{after} the applicable transition.
In the left part of ~\cref{fig:plug_and_final}, 
because the register has a pre-plug strand in $(a,\blank)$, it means that the instruction sublist $L_{(a,\blank)}$ has not been applied to it yet. Instruction sublists for the other transitions 
$(a,0) \to \ldots,\quad
(a,1) \to \ldots,\quad
(b,0) \to \ldots,\quad
(b,1) \to \ldots$ will be inert, 
not affecting the register. The first instruction of $L_{(a,\blank)}$ will remove this pre-plug strand so that any register in the $(a,\blank)$ configuration can be processed.
The instruction sublists will result in a configuration
like that of the bottom of~\cref{fig:transition-left-cell-has-1-first-half,fig:transition-left-cell-has-1-second-half}.
\cref{fig:transition-left-cell-has-1-second-half} shows instructions that affect the cell where the tape head \emph{was} (right cell), not where it will be \emph{next} (left cell),
which is why the left cell is the same in both~\cref{fig:transition-left-cell-has-1-first-half,fig:transition-left-cell-has-1-second-half}.
This almost represents the next Turing machine configuration,
but with the appropriate transition region covered by a \emph{post-plug strand}.
The final instruction in the entire list
(\cref{fig:plug_and_final})
removes this post-plug strand,
restoring the register configuration to be as shown in~\cref{fig:tm_example}.

The post-plug strand placed on the transition region at the end of simulating a transition has a different purpose from the pre-plug strand placed in instruction 1.
Its purpose is to prevent the register from updating its state multiple times in the same instruction iteration. For example, if a register has a post-plug strand on the $(b,1)$ region (indicating that it has been processed and that its next transition is $(b,1)$), and the instruction sublist that processes $(b,1)$ comes after, the register will be unaffected by $(b,1)$'s deprotecting instruction, keeping it inert throughout.
The final instruction in the entire iteration removes these post-plug strands from the registers, as seen in \cref{fig:plug_and_final}, preparing the registers for the next iteration of the instruction set.

Note the duality between pre-plug and post-plug instructions. \emph{All} pre-plug strands are included in the first instruction, though only one of them will bind (the one matching the applicable transition), and instruction strands removing \emph{all} post-plug strands are included as part of the last instruction, though only one will find its complementary post-plug strand to remove. On the other hand, each pre-plug instruction is removed more specifically, by adding a single complementary strand to remove it just prior to the sublist of instructions corresponding to the applicable transition. Similarly, each post-plug strand is added by itself, at the end of the instruction sublist corresponding to the applicable transition.

In the next section, we will describe the details of the instruction sublists that represent the Turing machine transitions.

\paragraph*{Sublist of instructions representing a single Turing machine transition}
\emph{Figure conventions.}
For the figures explaining SIMD||DNA instructions that simulate a single transition of the Turing machine
(\cref{fig:transition-left-cell-has-1-first-half} and beyond),
we use the following conventions in figures.
Several register configurations are shown,
but they are not necessarily consecutive.
Each is numbered with its absolute index in the list of all 86 instructions implementing the Turing machine of~\cref{fig:binary_tm}. An example with all instructions shown can be found in \cref{sec:appendix_full_overview}.
If two adjacent configurations have non-consecutive instruction indices,
this means that the instructions not shown are inert:
their strands do not affect the register in that configuration.
The instruction strands that have just been added are always shown above the register,
with a solid line if they will do a reaction as in~\cref{fig:simd-model},
and with a dashed line if that instruction strand is inert for that configuration.
The final configuration in each figure does not show any instruction strands,
but for all figures there is a followup figure showing what happens next from that configuration (possibly the followup is~\cref{fig:plug_and_final}, the final instruction in the entire program, removing the post-plug strand from the next applicable transition region).

Each Turing machine transition is individually processed by a sublist of instructions. The pre-plug strand is first removed by an instruction containing its complementary strand, so that its corresponding transition region in the cell can be used as a toehold, such as instruction 38 in~\cref{fig:transition-left-cell-has-1-first-half}. The next instructions then update the contents of the current cell to encode the symbol that the tape head writes. For example, in~\cref{fig:transition-left-cell-has-1-second-half}, the transition $(a,\blank) \to (b,\blank,L)$ is represented, and the strand encoding of \blank\ is placed in the right cell after the tape head writes on it and moves left. After that, the instructions check the contents of the tape head's new location and determine the Turing machine's next configuration. In~\cref{fig:transition-left-cell-has-1-first-half}, the tape head moves to the left cell and finds a 1, and the Turing machine goes to state $b$, so it leaves a post-plug strand on $(b,1)$'s transition region to show that the register has been processed for that instruction iteration, as seen in instruction 46.

\paragraph*{Left versus right tape head moves}
In the SIMD||DNA instructions implementing a single transition, 
there are two sublists: 
next-cell instructions and previous-cell instructions. As their names indicate, next-cell instructions update the contents of the tape head's destination, while previous-cell instructions update the contents of the tape head's former location. 
Other factors such as the symbol to be written on the current cell and the next applicable transition region only introduce minor variations in the instruction strands.

\begin{figure}[ht]
    \centering
    \includegraphics[width=12cm]{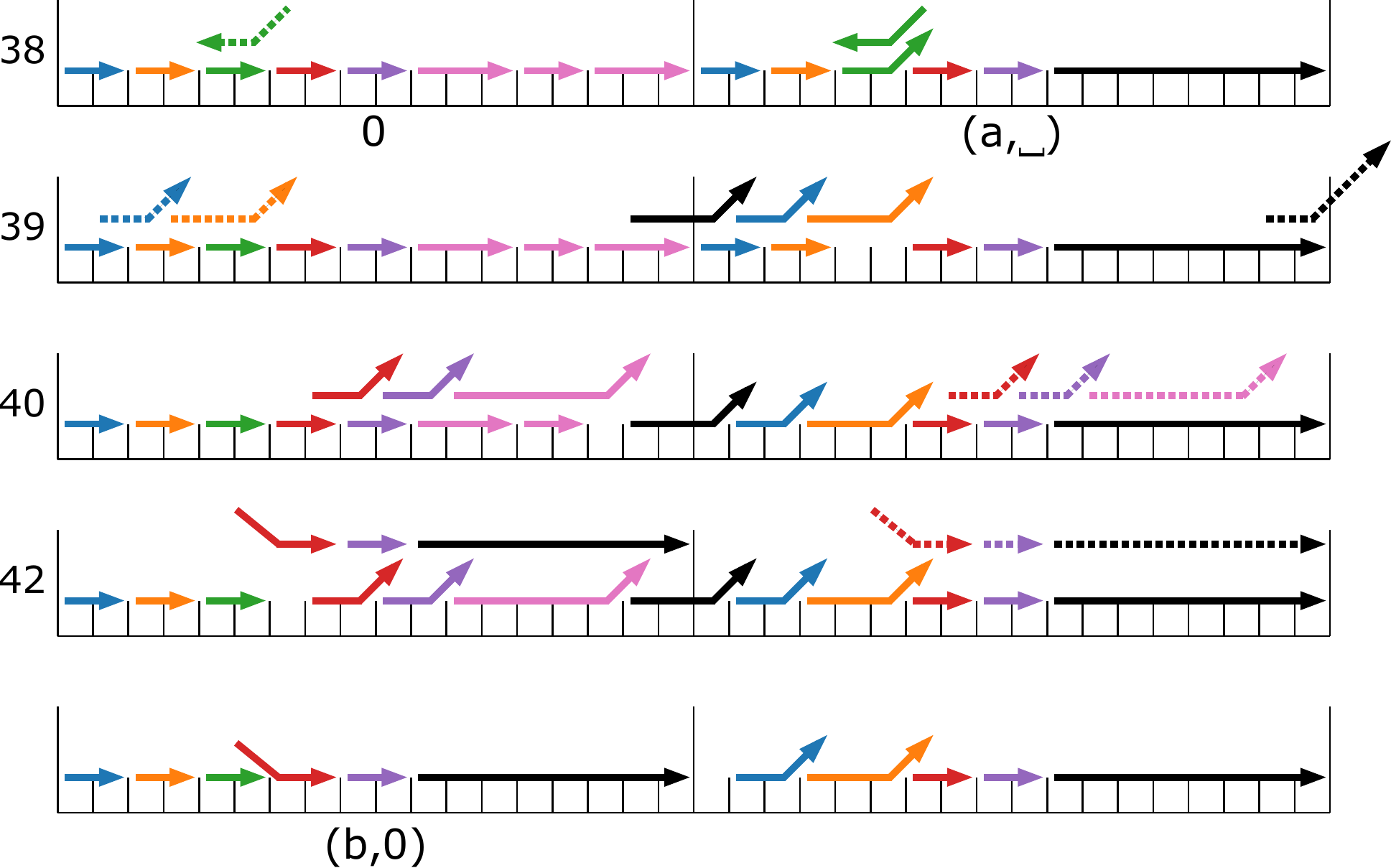}
    \caption{
    The next-cell instructions of the transition in \cref{fig:transition-left-cell-has-1-first-half} that are applicable when the cell to the left of the tape head has a 0.}
    \label{fig:transition-left-cell-has-0}
\end{figure}

\begin{figure}[ht]
    \centering
    \includegraphics[width=12cm]{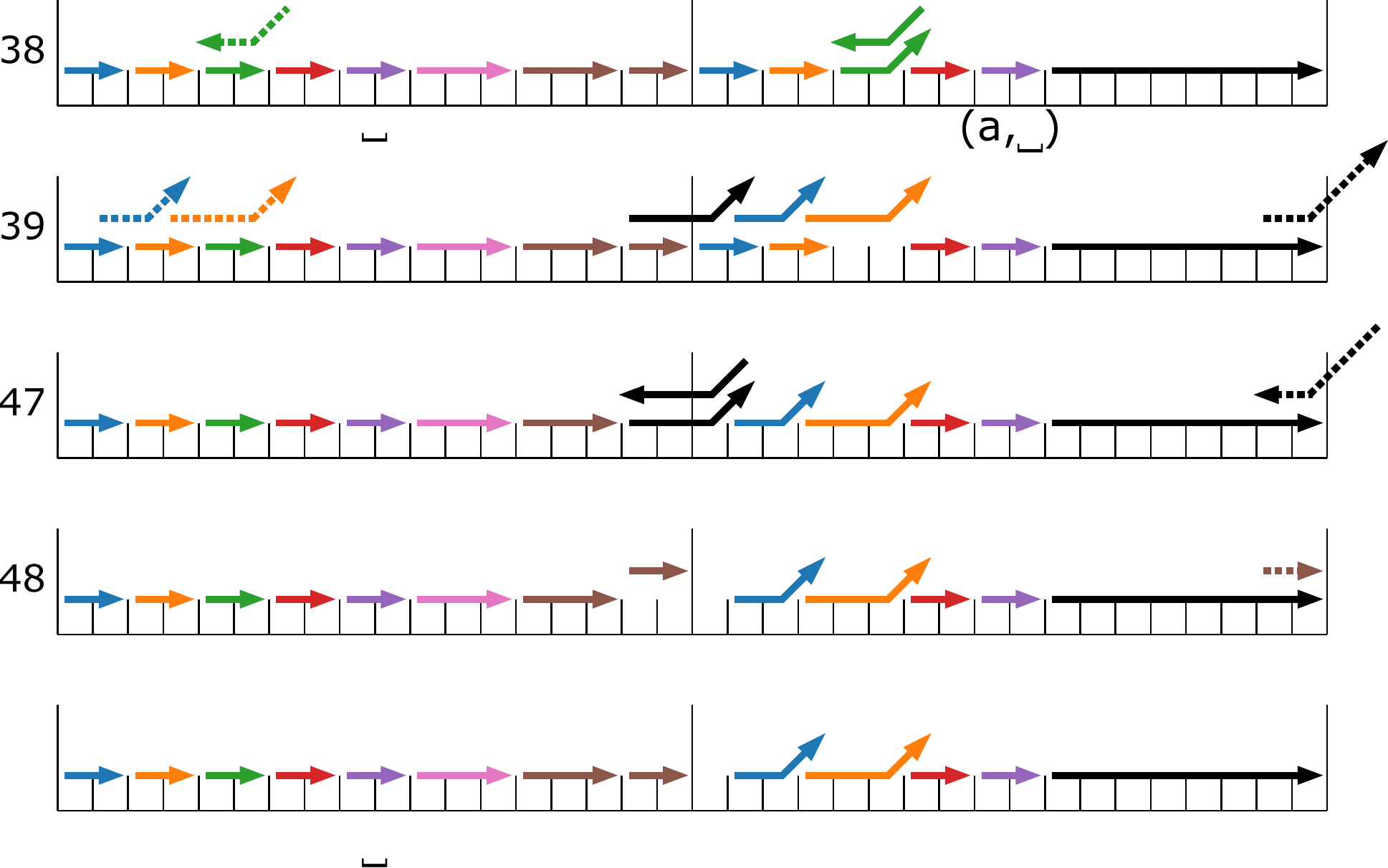}
    \caption{
    The next-cell instructions of the transition in \cref{fig:transition-left-cell-has-1-first-half} that are applicable when the cell to the left of the tape head has a \blank. 
    Because no transition exists for the state-symbol pair $(b,\text{\blank})$, the left cell is left unchanged.}
    \label{fig:transition-left-cell-has-blank}
\end{figure}

\begin{figure}[ht]
    \centering
    \includegraphics[width=12cm]{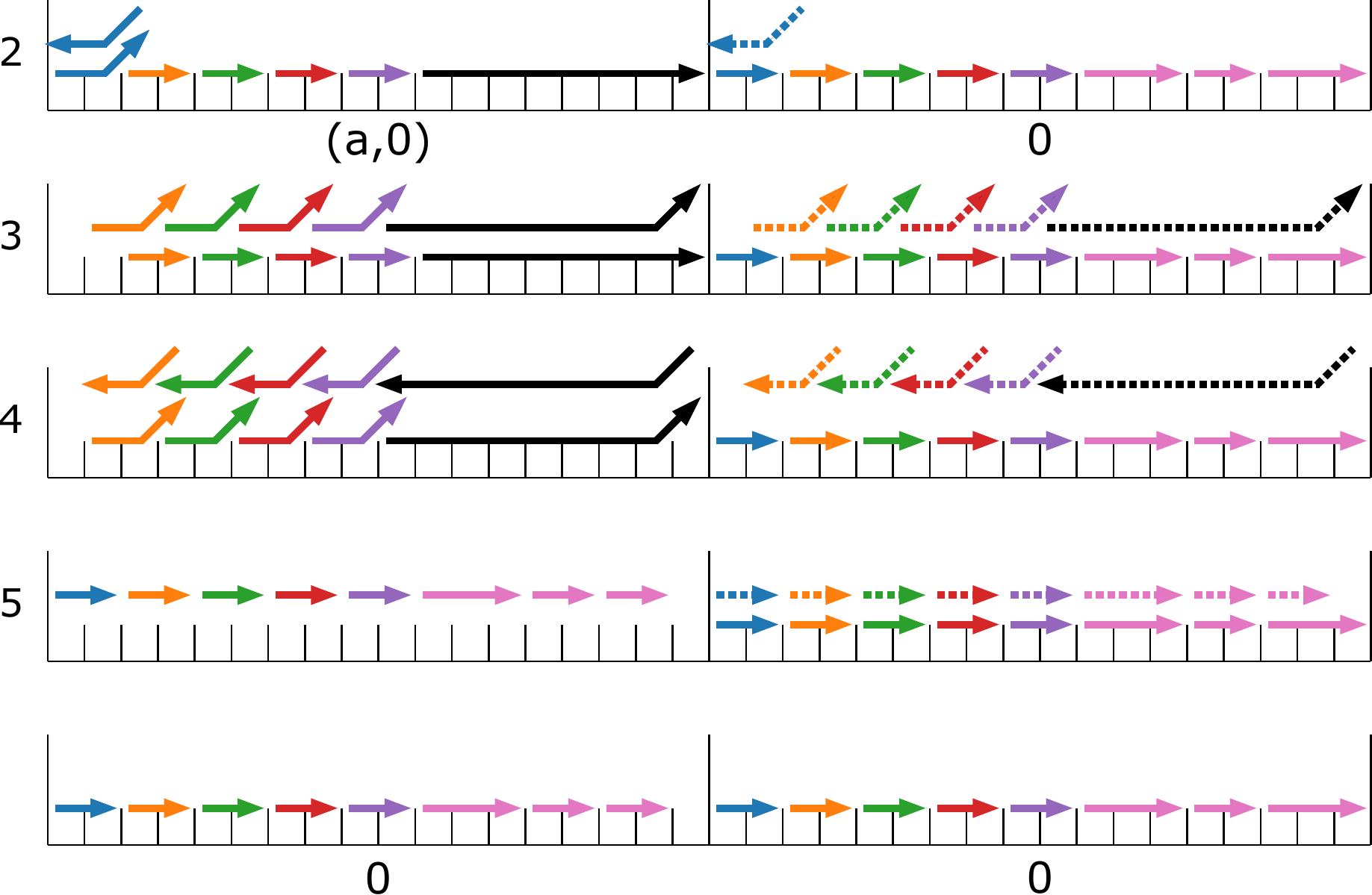}
    \caption{
    First half (previous-cell instructions) of transition $a,0 \to a,0,R$. Instruction 3's strands cascade to the right side of the current cell, while instruction 4 removes the previously introduced strands. Instruction 5 then covers up all the transition regions and adds the strands of the symbol to be written on the current cell (0 in this example), but leaving the rightmost domain exposed to act as a toehold for the next instructions.
    }
    \label{fig:transition-right-first-half}
\end{figure}

For transitions moving the tape head \emph{left}, the next-cell instructions precede the previous-cell instructions
(see~\cref{fig:transition-left-cell-has-1-first-half,fig:transition-left-cell-has-0,fig:transition-left-cell-has-blank,fig:transition-left-cell-has-1-second-half}).
For transitions moving the tape head \emph{right}, this order is reversed (see~\cref{fig:transition-right-first-half,fig:transition-right-cell-has-blank-second-half,fig:transition-right-cell-has-zero-second-half,fig:transition-right-cell-has-one-second-half}).

\paragraph*{Left tape head moves}
\cref{fig:transition-left-cell-has-1-first-half} shows the next-cell instructions for transition $(a,\blank) \to (b,\blank,L)$,
for the special case when the cell to the left of the tape head has the symbol 1. \cref{fig:transition-left-cell-has-0} shows the next-cell instructions for the same transition when the cell to the left of the tape head has the symbol 0, and \cref{fig:transition-left-cell-has-blank} shows the next-cell instructions when the cell to the left has the symbol \blank. 
Note that in any given configuration,
the same instructions will result in exactly one of the situations depicted in~\cref{fig:transition-left-cell-has-1-first-half,fig:transition-left-cell-has-0,fig:transition-left-cell-has-blank}.
Once these next-cell instructions are applied, the leftmost domain of the previous cell will serve as a toehold for the previous-cell instructions that follow in \cref{fig:transition-left-cell-has-1-second-half}.

\paragraph*{Right tape head moves}
For right transitions, the first three instructions are the previous-cell instructions, as shown in \cref{fig:transition-right-first-half}. The next-cell instructions follow, where \cref{fig:transition-right-cell-has-blank-second-half} shows what happens when the cell to the right of the tape head has a \blank, \cref{fig:transition-right-cell-has-zero-second-half} shows what happens when the cell to the right has a 0, and \cref{fig:transition-right-cell-has-one-second-half} shows what happens when the cell to the right has a 1.
As with the left-cell instructions, the next-cell situations depicted in~\cref{fig:transition-right-cell-has-blank-second-half,fig:transition-right-cell-has-zero-second-half,fig:transition-right-cell-has-one-second-half} are disjoint.

\paragraph*{Final deprotecting instruction}

\cref{fig:plug_and_final} shows the final deprotecting instruction,
which removes the post-plug strand put in place during the last instruction of the non-inert instruction sublists.
This puts the register back into a ``clean'' configuration representing a Turing machine configuration,
such as those shown in~\cref{fig:tm_example},
opening up new toeholds for the next iteration of instructions.

\subsection{Complexity of construction}
A common metric of ``complexity'' of DNA systems is the number of unique domains they require.
Fewer is better because it is a nontrivial task to design \emph{orthogonal} domains:
domains that, if they are not perfectly complementary,
will have low binding affinity.
Low domain complexity is particular important in SIMD||DNA,
where each domain is considered ``toehold-length'':
sufficiently short (5-7 bases)
that the off-rate of a strand bound by a single domain is large enough to detach in a short amount of time.
There are only $4^7 = 16384$ DNA sequences of length $7$.
In practice even fewer are available:
half are complementary to the other half, leaving only $8192$ available to assign to the unstarred versions of each domain.
DNA sequence design heuristics such as the ``3-letter code'' (using only A,T,C for forward strands, thus only A,T,G for the register and reverse strands)
reduce this number to $3^7 = 2187$.
Reasonable design constraints,
e.g., avoiding almost equal domains such as $5'$-AAAAAAG-$3'$ and $5'$-AAAAAAA-$3'$,
which both bind almost equally strongly to $3'$-TTTTTTT-$5'$,
further limit the set of available domains.

Our construction uses $d = 2t + 8$ total unique DNA domains (which repeat throughout the register), where $t$ is the number of transitions of the simulated Turing machine. Each transition is represented by 2 domains,
assumed to be bound strongly enough not to spontaneously dissociate. 
(The construction can be altered to use more domains, in case this assumption is overly ideal.)
To simulate a Turing machine with space bound $s$,
the register has $s$ ``cells'',
where each cell is simply a copy of one each of the $d$ domains $1,\ldots d$.
Thus, if each domain consists of $k$ nucleotides,
the register strand has $k \cdot s \cdot (2t + 8)$ total nucleotides.
There is an 11-state, 3-symbol universal Turing machine (directly simulating another Turing machine) with 32 transitions~\cite{neary2006small},
giving $2 \cdot 32 + 8 = 72$ total domains required in the worst case.
However, specialized non-universal Turing machines with a smaller number of transitions (for example the 5-transition binary incrementor of~\cref{fig:tm_example}) could accomplish many computationally sophisticated tasks.

Each Turing machine transition can be represented by approximately 16 SIMD||DNA instructions. The exact number varies depending on the specific properties of the transition in question, such as the the direction of the tape head, the symbol to be written, and whether any of the possible next configurations are halting or not. This range has constant upper and lower bounds, however, so the total number of instructions is in $O(t)$, where $t$ is the number of transitions in the Turing machine. Because the size of the cell increases in $O(t)$ due to the transition regions, the number of DNA strands present in some instructions also scales by a factor of $O(t)$, most notably the instructions that cause a cascade of toehold exchanges, such as instructions 3 to 7 in  \cref{fig:transition-right-first-half,fig:transition-right-cell-has-blank-second-half}.
\section{Conclusion}
\label{sec:conclusion}

Our construction,
like the Rule 110 simulation of Wang, Chalk, and Soloveichik~\cite{wang2019simd},
is not Turing universal because it simulates a \emph{space-bounded} Turing machine.
Truly universal computation should be possible without advanced knowledge of the space requirement.
An interesting question
(raised also in~\cite{wang2019simd})
is whether a suitable augmentation of the SIMD||DNA model could allow Turing-universal computation.
This would require unbounded polymers such as those used in the two-stack machine DNA implementation of Qian, Soloveichik, and Winfree~\cite{qian2010efficient}.
That paper showed Turing universal computation in the case where only a single copy of certain strands are permitted to exist in solution, simulating only a single stack machine at a time,
in contrast to the SIMD||DNA model,
where we can operate on many registers, each representing their own Turing machine, in parallel.

Technically the strand displacement reactions of the SIMD||DNA model are currently powerful enough to grow arbitrarily large polymers from a fixed set of strands,
as in~\cite{qian2010efficient},
by alternating top and bottom strands,
making a long double-helix with nicks on the top \emph{and} bottom.
However, it is difficult to see how to use the ability of SIMD||DNA to exploit this to simulate a Turing machine represented in this way.
If there are multiple bottom strands,
i.e., there is a nick on the bottom,
then any strand displacement of top strands,
upon reaching this nick,
would separate the polymer into two complexes to the left and right of this nick,
and the right polymer would be lost in the wash step.
One could imagine, however, augmenting the model to allow,
for example,
3-arm junctions,
which could be used to do strand displacement that crosses over the boundary between two bottom strands without separating them (since they would be joined to each other by a strong domain representing the third arm ``below'' the main helix).

Although some of the toehold exchanges in the SIMD||DNA model are reversible based on the principles of DNA strand displacement, we make the assumption that the applied instructions are not undone by the displaced strands. This is based on the assumption that the instruction strands are present in a sufficiently high concentration that reversal is unlikely. Because multiple registers can be present in the same solution, another possibility to consider is a displaced DNA strand from Register A binding to an open toehold in Register B, such that the attachment is irreversible even with the presence of a high concentration of instruction strands. One open question is to design a system that factors in these possibilities, reducing the likelihood of unexpected strand displacement results.

Another open question is whether a more domain-efficient encoding exists for the Turing machine construction. Given $n$ transitions, $2n$ domains are required to represent them, which has $O(n)$ complexity. However, given $d$ domains, there are $2^{d-1}$ possible nick patterns among the attached strands, which makes  $O(\log n)$ domain complexity possible in theory.

\clearpage
\bibliography{ref}

\clearpage
\appendix
\section{Appendix: Next-cell Instructions}
\label{sec:appendix_next_cell}
\begin{figure}[ht!]
    \centering
    \includegraphics[width=9cm]{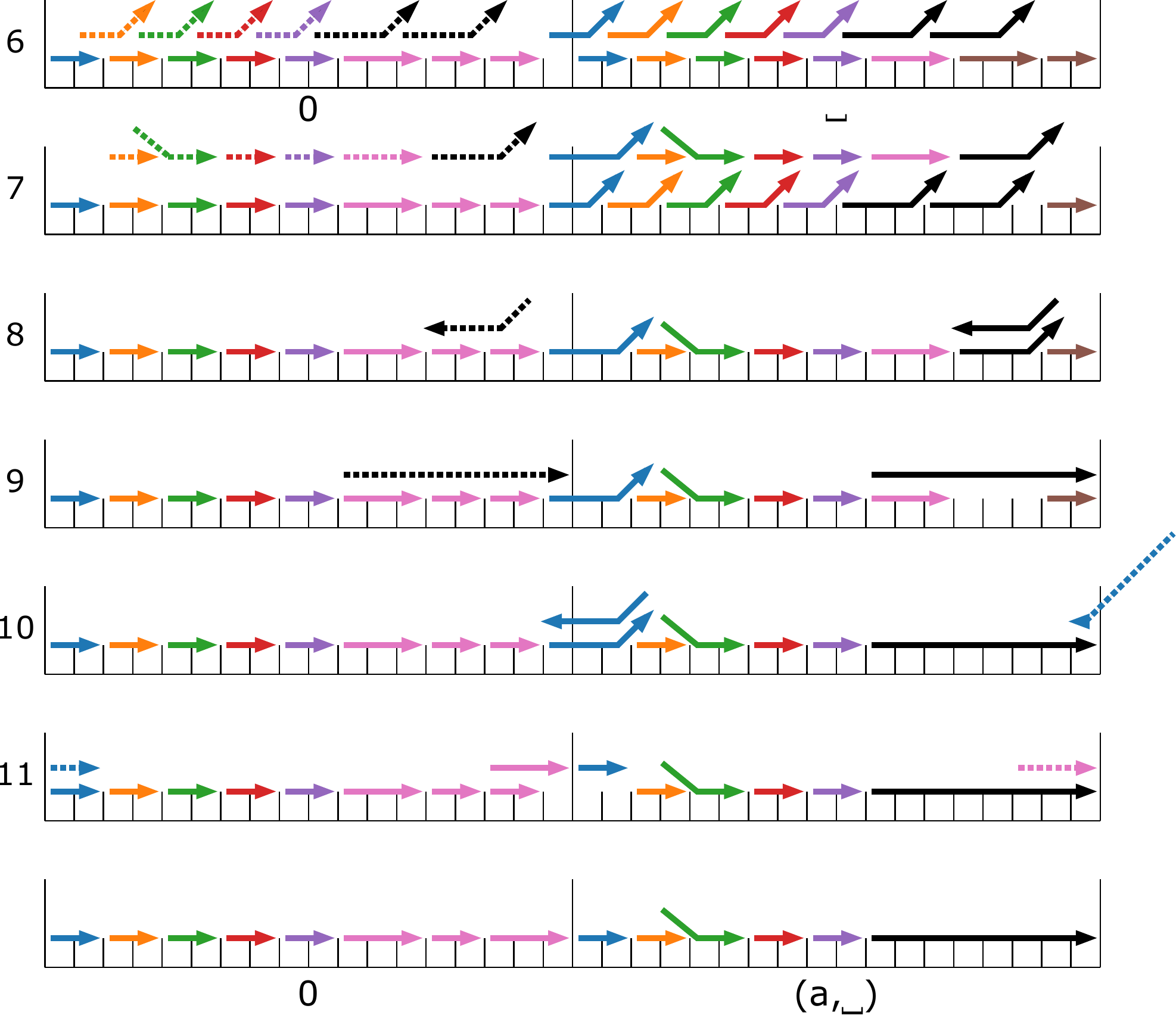}
    \caption{The next-cell instructions for when the right cell contains a blank. The Turing machine enters the $(a,\text{\blank})$ configuration afterward.}
    \label{fig:transition-right-cell-has-blank-second-half}
\end{figure}

\begin{figure}[ht!]
    \centering
    \includegraphics[width=9cm]{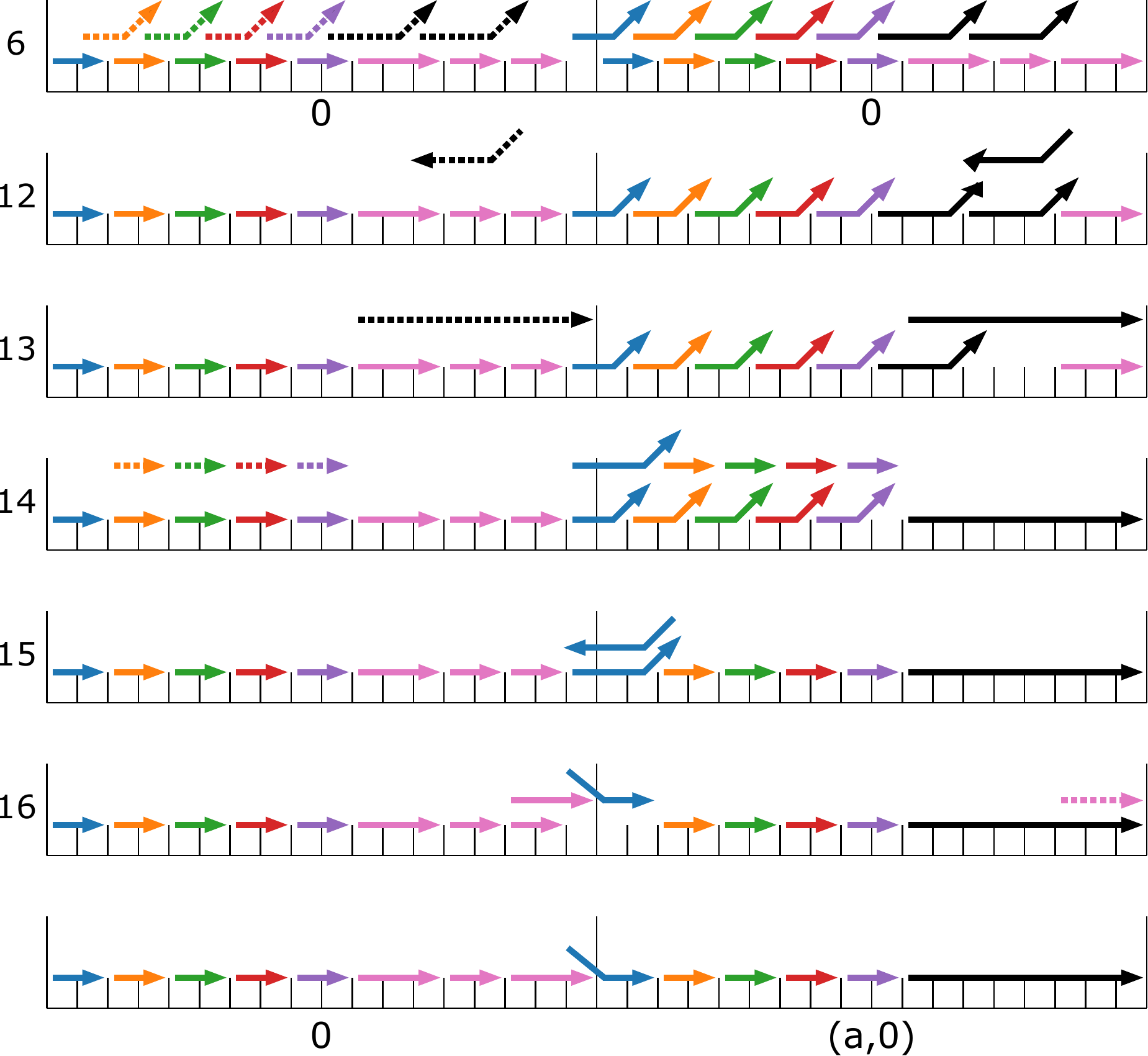}
    \caption{The next-cell instructions for when the right cell contains the symbol 0, entering the $(a,0)$ configuration.}
    \label{fig:transition-right-cell-has-zero-second-half}
\end{figure}

\begin{figure}[ht!]
    \centering
    \includegraphics[width=9cm]{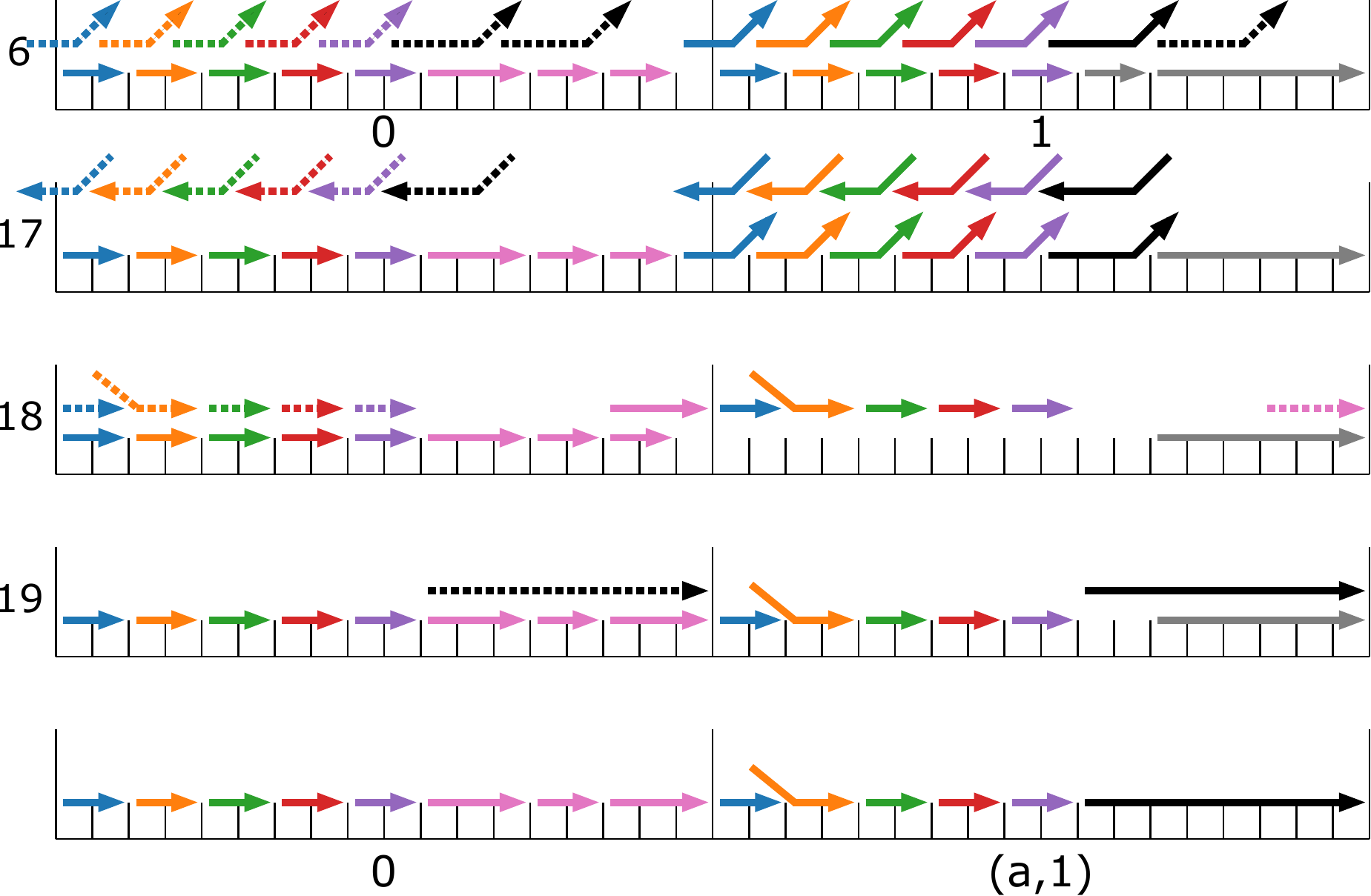}
    \caption{The next-cell instructions for when the right cell contains the symbol 1, entering the $(a,1)$ configuration.}
    \label{fig:transition-right-cell-has-one-second-half}
\end{figure}

\clearpage

\section{Appendix: $(a, 1)$ Transition Overview}
\label{sec:appendix_a1}

\begin{figure}[ht!]
    \centering
    \includegraphics[width=\textwidth]{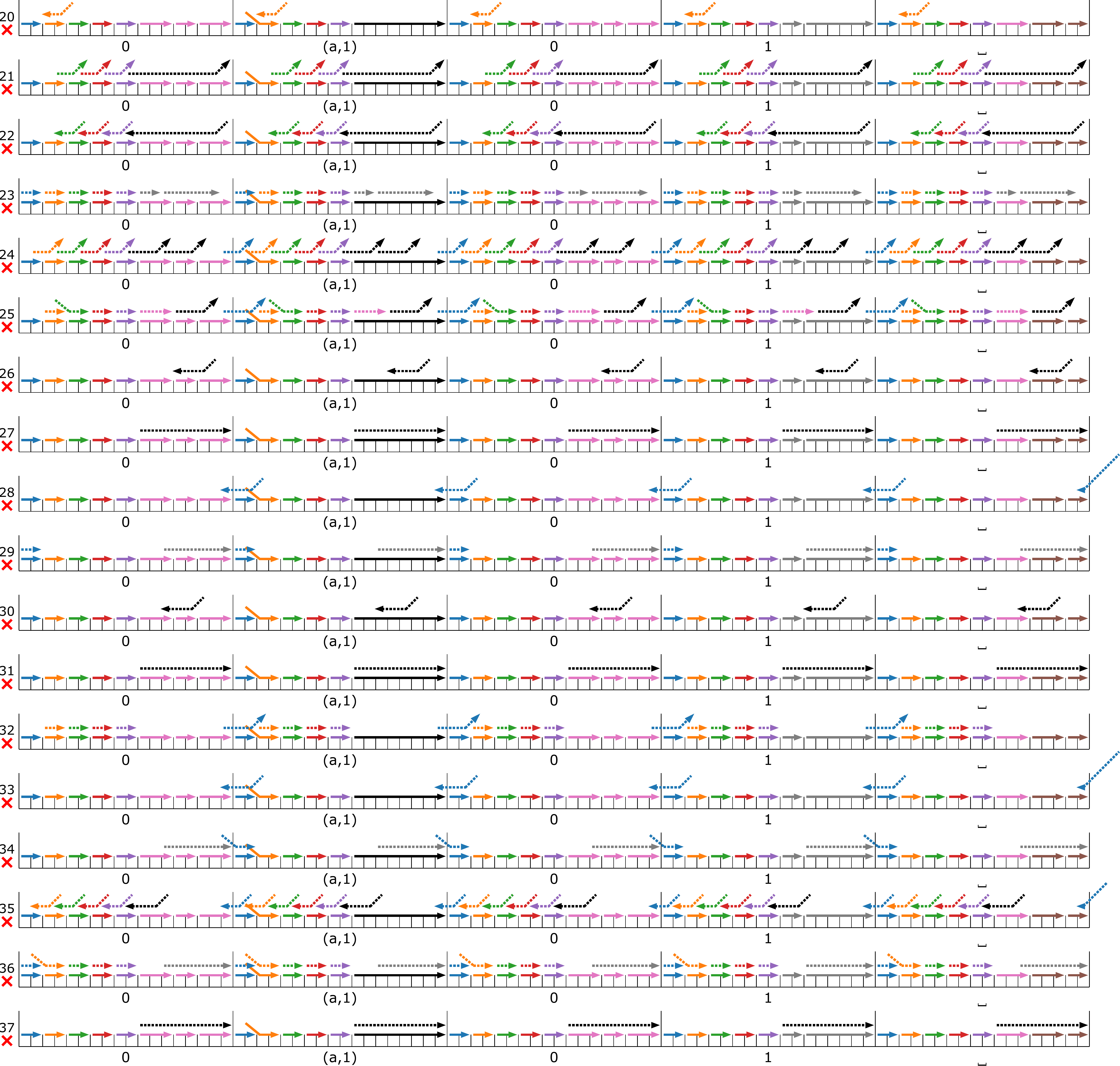}
    \caption{
    An overview of instructions 20 to 37 of the construction, which represent the $(a, 1) \to (a,1,R)$ transition of the Turing machine shown in \cref{fig:binary_tm}. Because the register in this example is not in the $(a, 1)$ configuration, the instructions in this section do not affect the register, as no toehold is present.}
    \label{fig:a1_instructions}
\end{figure}

\newpage

\section{Appendix: Full Instruction Overview}
\label{sec:appendix_full_overview}

\begin{figure}[ht]
    \centering
    \includegraphics[width=0.65\textwidth]{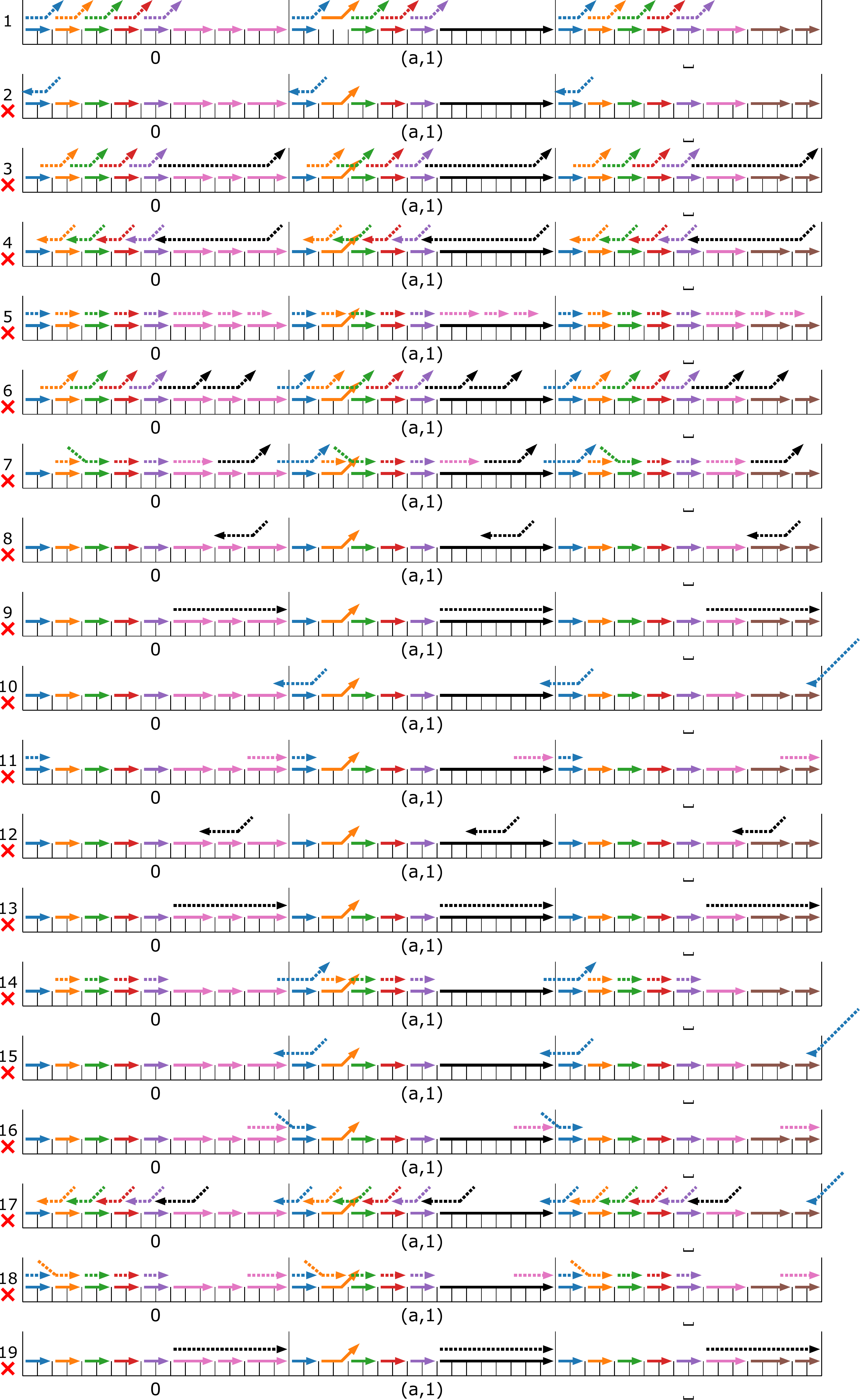}
    \caption{The initial pre-plug instruction, as well as the $(a,0)$ instruction sublist, for a register whose next transition is $(a,1)$. Note all instructions after the first are inert, because the pre-plug strand blocks all subsequent instructions (as intended since $(a,0)$ is not the applicable transition.)}
    \label{fig:appendix_a0-2}
\end{figure}

\begin{figure}[ht]
    \centering
    \includegraphics[width=\textwidth]{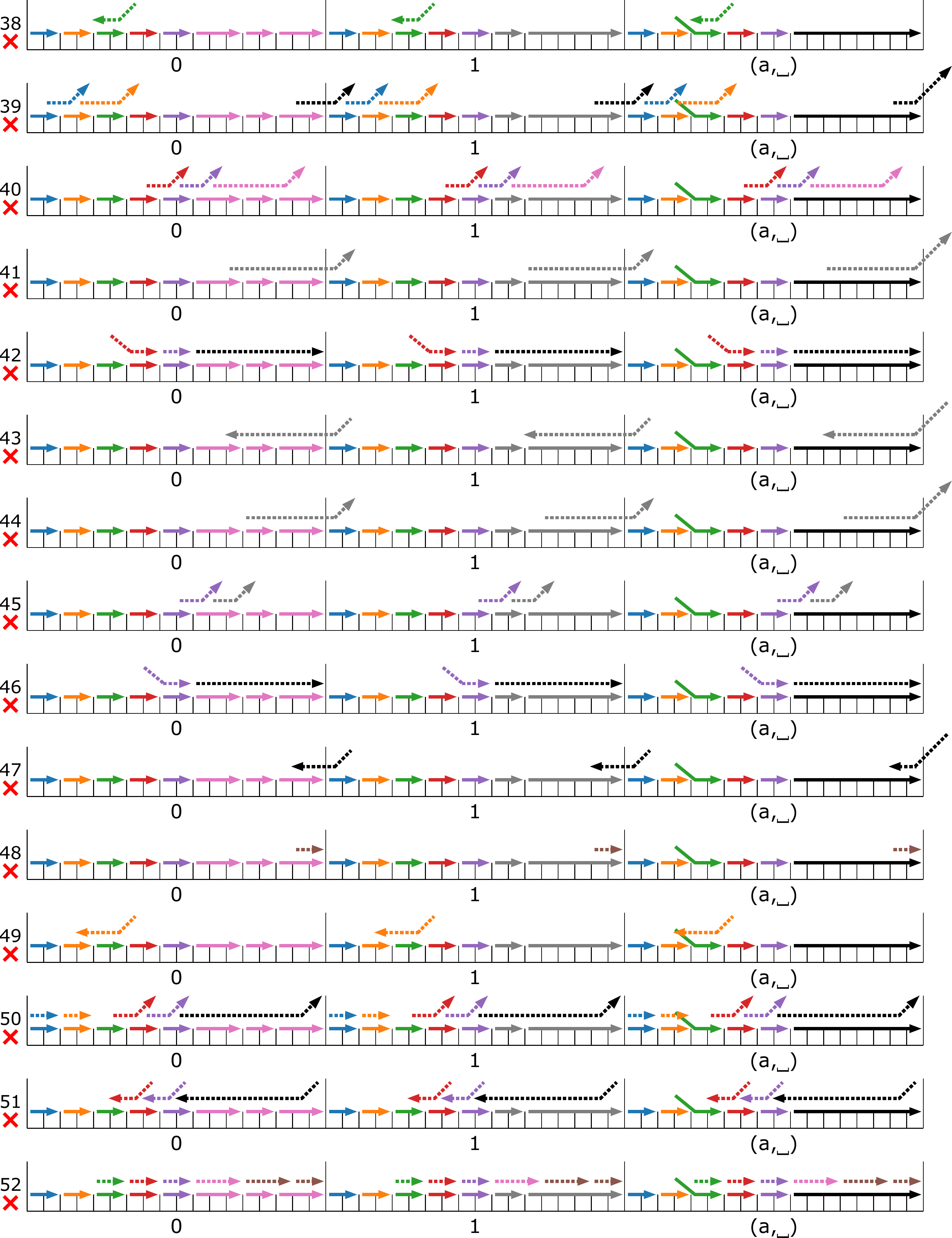}
    \caption{The $(a,\blank)$ instruction sublist after the register was processed by the $(a,1)$ instruction sublist. Even though the register is now in the $(a,\blank)$ configuration, the post-plug strand introduced in instruction 25 prevents the register from being updated a second time using the $(a,\blank)$ instructions.}
    \label{fig:appendix_a_-2}
\end{figure}

\begin{figure}[ht]
    \centering
    \includegraphics[width=\textwidth]{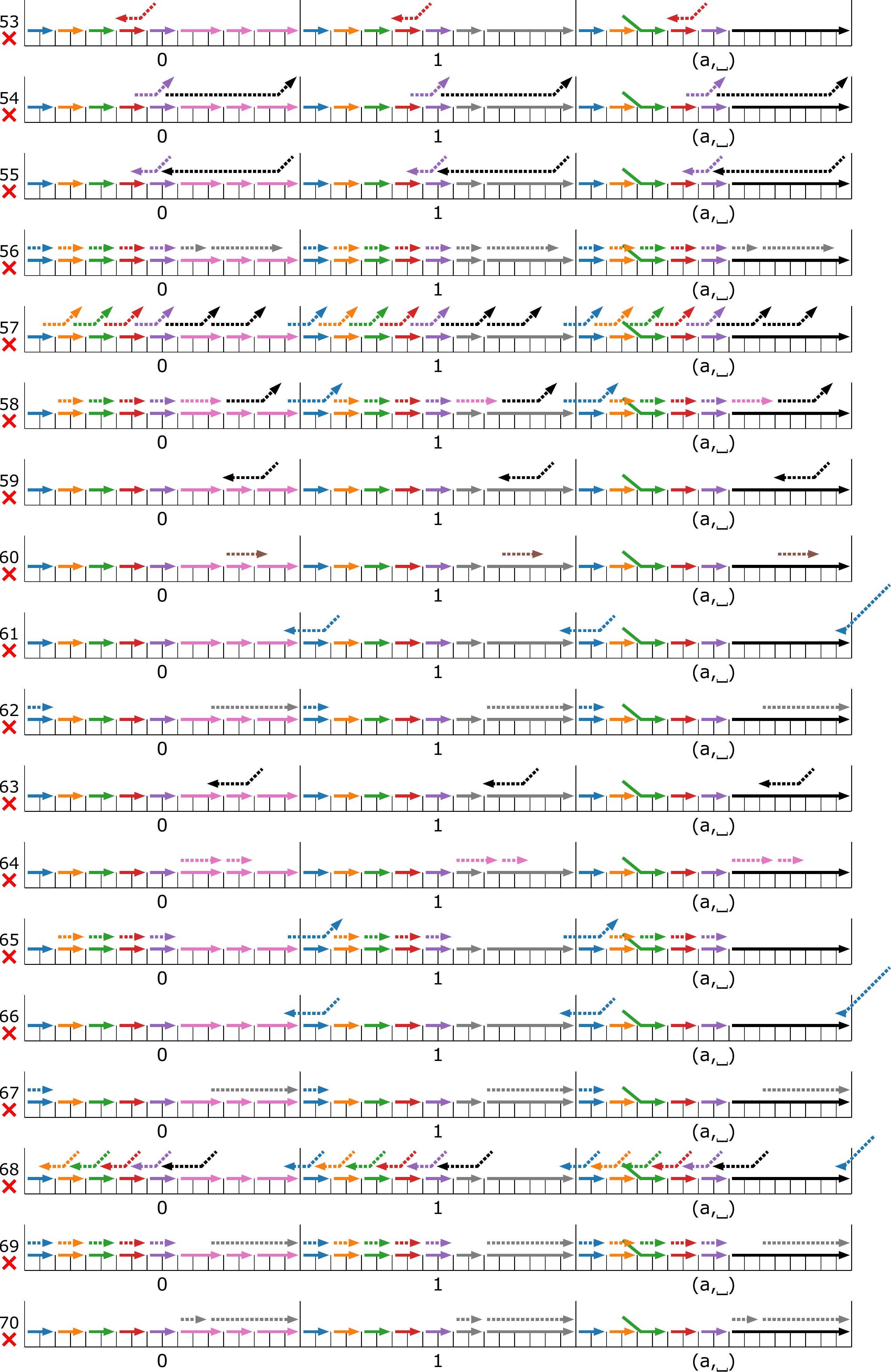}
    \caption{The $(b,0)$ instruction sublist after the register was processed by the $(a,1)$ instruction sublist.}
    \label{fig:appendix_b0-2}
\end{figure}

\begin{figure}[ht]
    \centering
    \includegraphics[width=\textwidth]{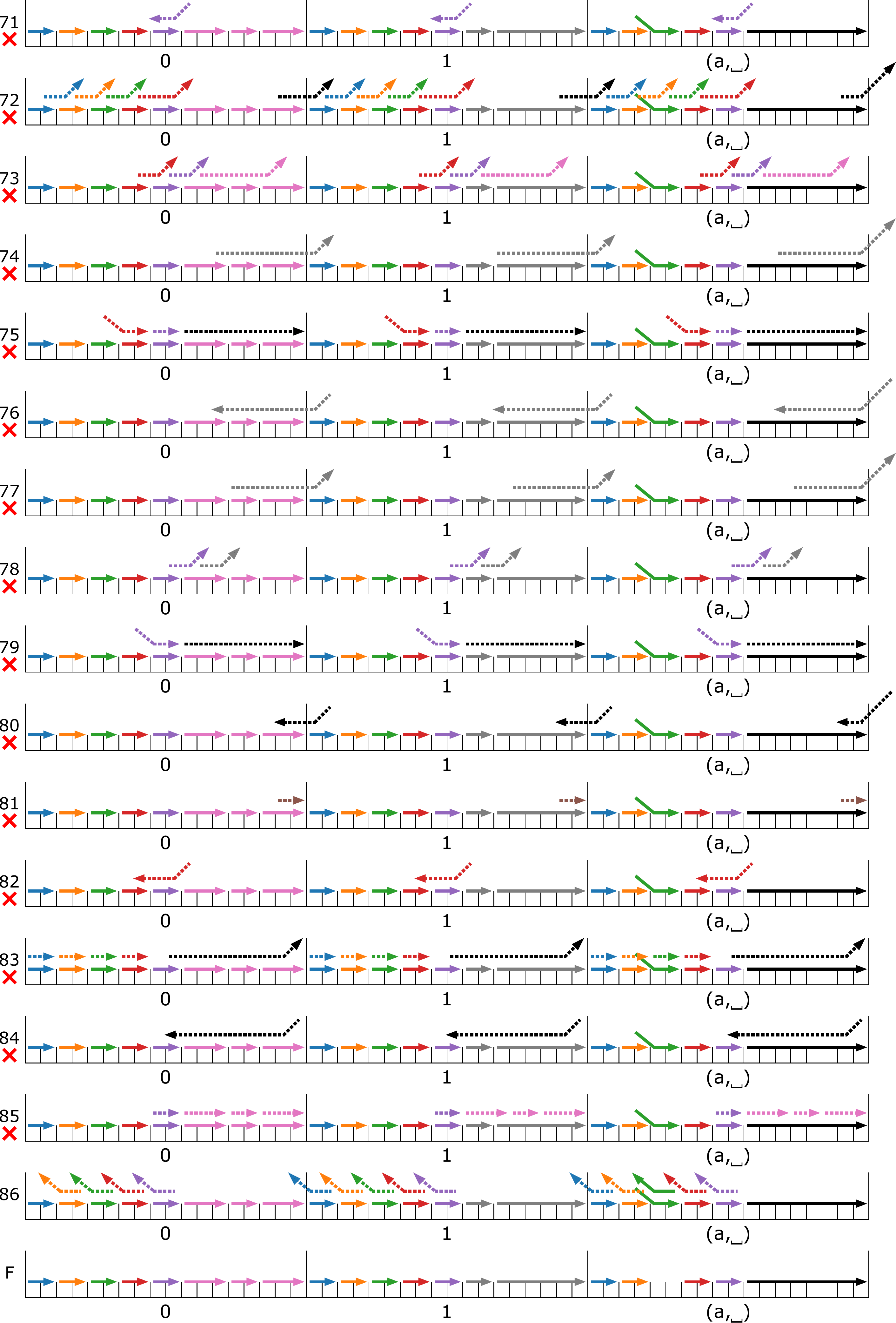}
    \caption{The $(b,1)$ instruction sublist after the register was processed by the $(a,1)$ instruction sublist, as well as the final deprotecting instruction that removes the post-plug strand.}
    \label{fig:appendix_b1-2}
\end{figure}

The following figures show the full instruction set on a register whose current value is 01\blank, where the tape head is on 1 and the current state is $a$. Because the $(a,1)$ transition is represented in this configuration, the instruction sublists that represent the other transitions (\cref{fig:appendix_a0-2,fig:appendix_a_-2,fig:appendix_b0-2,fig:appendix_b1-2}) are fully inert.

\begin{figure}[ht]
    \centering
    \includegraphics[width=\textwidth]{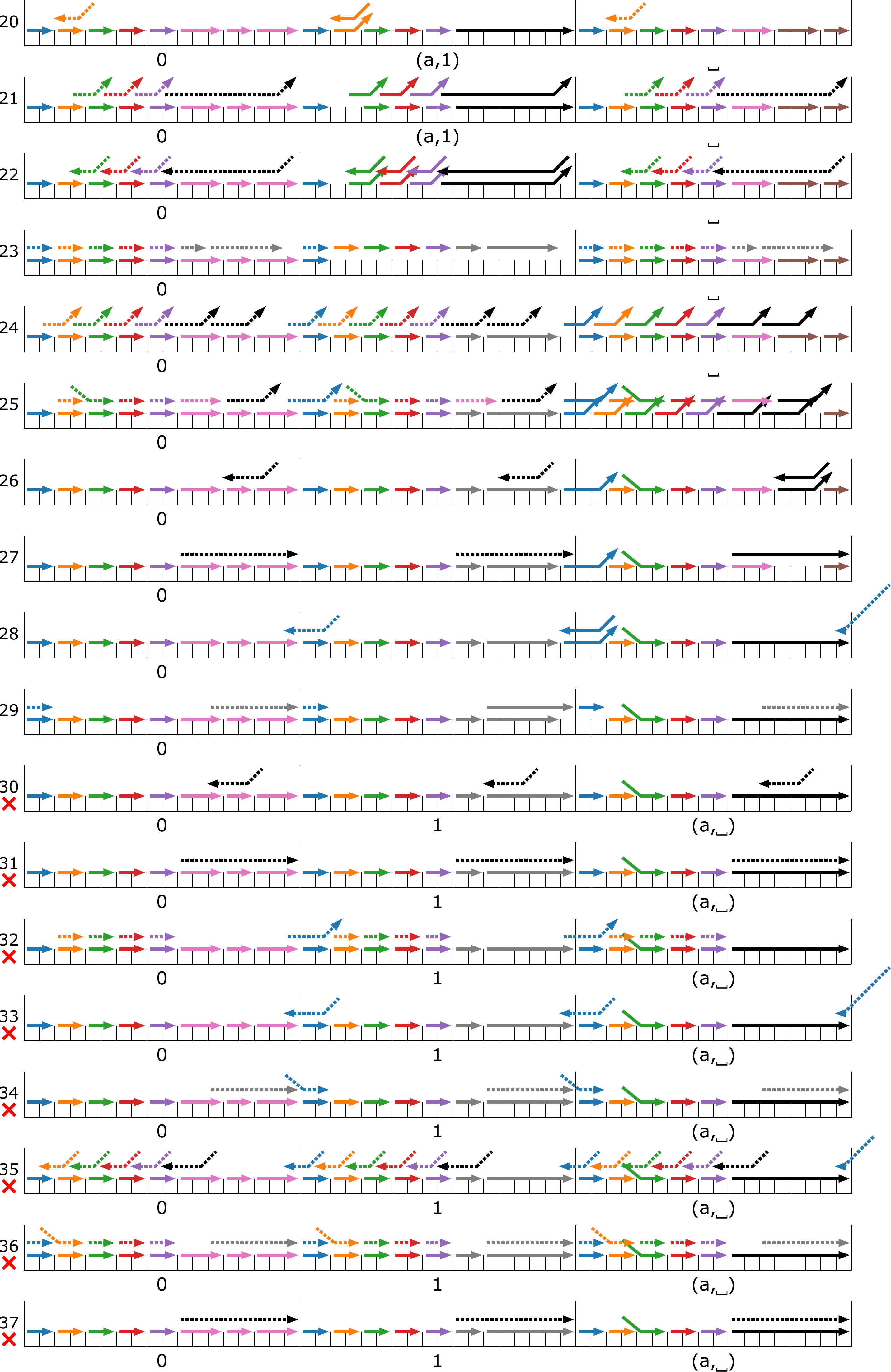}
    \caption{The $(a,1)$ instruction sublist for a register whose next transition is $(a,1)$. Instructions 30 to 37 are inert, because they apply to the cases when the cell to the right of $(a,1)$ contains a 0 or a 1.}
    \label{fig:appendix_a1-2}
\end{figure}

\end{document}